\documentclass{emulateapj}
\usepackage{apjfonts}
\usepackage{lscape}

\newcommand{\Lp}{L_{\rm peak}}

\newcommand{\Lbol}{L_{\rm bol}}

\newcommand{\nh}{N_{\rm H}}
\newcommand{\nhi}{N_{\rm H\,I}}

\newcommand{\Mdot}{\dot{M}}

\newcommand{\etal}{et al.}

\newcommand{\mlk}{M/L_{\rm K}}
\newcommand{\mlkfrac}{\frac{M}{L_{\rm K}}}
\newcommand{\mbul}{M_{\ast}}
\newcommand{\vvir}{V_{\rm vir}}

\newcommand{\qeos}{q_{\rm EOS}}
\newcommand{\fgas}{f_{\rm gas}}
\newcommand{\zgal}{z_{\rm gal}}
\newcommand{\mbh}{M_{\rm BH}}

\newcommand{\nLP}{\dot{n}(M_{\rm BH})}
\newcommand{\nLp}{\nLP}

\newcommand{\reducechi}{\chi^{2}/\nu}
\newcommand{\mtot}{M_{\ast,\, {\rm tot}}}
\newcommand{\mnew}{M_{\ast,\, {\rm new}}}

\newcommand{\muv}{M_{280}}
\newcommand{\luv}{L_{280}}
\newcommand{\suv}{\sigma_{280}}
\newcommand{\mluv}{M/L_{280}}

\newcommand{\EV}[1]{\langle #1 \rangle}
\newcommand{\mluveff}{\EV{\mluv}}

\newcommand{\tmerge}{t_{\rm merge}}
\newcommand{\tmerger}{\tmerge}
\newcommand{\sfr}{\dot{M_{\ast}}}
\newcommand{\tsfr}{t_{\ast}}
\newcommand{\scalesinglefig}{\epsscale{1.15}}

\shorttitle{Merging Galaxy and Quasar Luminosity Functions}
\shortauthors{Hopkins \etal}
\slugcomment{Submitted to ApJ, February 13, 2006}
\begin{document}

\title{The Relation Between Quasar and Merging Galaxy Luminosity Functions
and the Merger-Driven Star Formation History of the Universe}
%\title{Linking the Observed Populations of Quasars and Merging Galaxies}
\author{Philip F. Hopkins\altaffilmark{1}, 
Rachel S. Somerville\altaffilmark{2}, 
Lars Hernquist\altaffilmark{1}, 
Thomas J. Cox\altaffilmark{1}, 
Brant Robertson\altaffilmark{1}, 
\&\ Yuexing Li\altaffilmark{1}
}
\altaffiltext{1}{Harvard-Smithsonian Center for Astrophysics, 
60 Garden Street, Cambridge, MA 02138}
\altaffiltext{2}{Max-Planck-Institut f\"{u}r Astronomie, 
K\"{o}nigstuhl 17, Heidelberg 69117, Germany}

\begin{abstract}

Using a model for the self-regulated growth of supermassive black
holes in mergers involving gas-rich galaxies, we study the
relationship between quasars and the population of merging galaxies
and thereby predict the merger-driven star formation rate density of
the Universe.  In our picture, mergers drive gas inflows, fueling
nuclear starbursts and ``buried'' quasars until feedback disperses the
gas, allowing the quasar to be briefly visible as a bright optical
source.  As black hole accretion declines, the quasar dies and the
stellar remnant relaxes passively with properties and 
correlations typical of red, elliptical galaxies.  By simulating the
evolution of such events, we demonstrate that the observed {\em
statistics} of merger rates/fractions, luminosity functions, mass
functions, star formation rate distributions, quasar luminosity
functions, quasar host galaxy luminosity functions, and elliptical/red
galaxy luminosity and mass functions are self-consistent.  We use our
simulations to de-convolve both the quasar and merging galaxy
luminosity functions to determine the birthrate of black holes of a
given final mass and merger rates as a function of the total stellar
mass and the mass of new stars formed during a merger.  From this, we
predict the merging galaxy luminosity function in various observed
wavebands (e.g.\ UV, optical, and near-IR), color-magnitude relations,
mass functions, absolute and specific star formation rate
distributions and star formation rate density, and quasar host galaxy
luminosity function, as a function of redshift from $z=0-6$.  We
invert this relationship to predict e.g.\ quasar luminosity functions
from observed merger luminosity functions or star formation rate
distributions.  Our results show good agreement with observations, but
idealized models of quasar lightcurves give inaccurate estimates and
are ruled out by comparison of merging galaxy and quasar observations
at $>99.9\%$ confidence, provided that quasars are triggered in
mergers.  Using only observations of quasars, we estimate the
contribution of mergers to the star formation rate density of the
Universe to high redshifts, $z\sim4$, and constrain the evolution in the
characteristic initial gas fractions of quasar and spheroid-producing
mergers.

\end{abstract}


\section{Introduction}
\label{sec:intro}

It is believed that our Universe evolved hierarchically, with small
objects merging to form progressively larger systems.  Observations of
interacting galaxies support this view and led Toomre \& Toomre (1972)
and Toomre (1977) to propose the ``merger hypothesis,'' according to
which ellipticals originate when spiral galaxies collide (for reviews,
see e.g.\ Barnes \& Hernquist 1992; Barnes 1998; Schweizer 1998).  It
has also been established that supermassive black holes are ubiquitous
in the centers of galaxies (e.g. Kormendy \& Richstone 1995; Richstone
et al.\ 1998; Kormendy \& Gebhardt 2001) and that their masses
correlate with either the masses (Magorrian et al. 1998) or the
velocity dispersions (i.e. the $M_{\rm BH}$-$\sigma$ relation:
Ferrarese \& Merritt 2000; Gebhardt et al. 2000) of their host
spheroids.  These relations argue that supermassive black holes and
galaxies are linked in an evolutionary manner, as implied by
simulations showing that self-regulated black hole growth in a galaxy
merger has a significant impact on the structure of the remnant (Di
Matteo et al. 2005).

Observations also indicate that galaxy mergers produce starbursts and
fuel black hole growth, and that both play roles in structuring
ellipticals.  Infrared luminous galaxies appear to be powered at least
in part by nuclear starbursts (e.g.\ Soifer et al. 1984a,b; Sanders et
al. 1986, 1988a,b; for a review, see e.g.\ Soifer et al. 1987), and
the most intense of these, ultraluminous infrared galaxies (ULIRGs),
are always associated with mergers (e.g.\ Allen et al. 1985; Joseph \&
Wright 1985; Armus et al. 1987; Kleinmann et al. 1988; Melnick \&
Mirabel 1990; for reviews, see Sanders \& Mirabel 1996 and Jogee
2005).  Radio observations reveal large quantities of dense gas in the
centers of ULIRGs (e.g.\ Scoville et al. 1986; Sargent et al. 1987,
1989), providing material to feed black hole growth and to boost the
concentration and central phase space density to match those of
ellipticals (see, e.g.\ Carlberg 1986; Gunn 1987; Lake 1989; Hernquist
1992, 1993a; Hernquist et al.\ 1993; Robertson et al.\ 2006a).  Some
ULIRGs have ``warm'' IR spectral energy distributions (SEDs),
suggesting that they harbor buried quasars (e.g.\ Sanders et
al. 1988c), as implied by X-ray observations of growing black holes in
NGC6240 (Komossa et al.\ 2003) and other ULIRGs, which are
heavily obscured at optical and infrared wavelengths (e.g.  Gerssen et
al.\ 2004; Max et al.\ 2005; Alexander et al.\ 2005a,b; Borys et al.\
2005).

These lines of evidence, and the overlap between the bolometric
luminosities of ULIRGs and quasars, suggest that quasars originate
from an IR luminous phase of galaxy evolution caused by mergers
(Sanders et al. 1988a).  This proposal is supported by observations of
nearby hosts of bright quasars which reveal features characteristic of
mergers (e.g.\ Stockton 1978; Heckman et al. 1984; Stockton \&
MacKenty 1987; Stockton \& Ridgway 1991; Hutchings \& Neff 1992;
Bahcall et al. 1994, 1995, 1997; Canalizo \& Stockton 2001) and
starbursts
\citep[e.g.,][]{Brotherton99,Brotherton02,Canalizo00,Yip04,VandenBerk05}.

Observations of individual merging systems and merger remnants
\citep[e.g.,][]{LakeDressler86,Doyon94,Oliva95,
ShierFischer98,James99} have shown that their kinematic and
photometric properties, including velocity dispersions,
concentrations, stellar masses, light profiles, and phase space
densities, are consistent with their eventual evolution into typical
$\sim L_{\ast}$ elliptical galaxies.  Furthermore, the correlations
obeyed by these mergers and remnants
\citep[e.g.,][]{Genzel01,RJ04,RJ05} are similar to e.g.\ the observed
fundamental plane and Kormendy relations for relaxed ellipticals, and
consistent with evolution onto these relations as their stellar populations
age.

While these studies suggest that nearby gas-rich mergers pass through
starburst and quasar phases and evolve into ellipticals, the implications
of these findings at higher redshifts are unclear and little is known
about the {\em statistics} of galaxy mergers in this context.  In particular, it is
difficult to measure the detailed properties of distant quasar hosts
\citep[e.g.,][]{Schweizer82,Dunlop03,Floyd04}, and it is not clear if
there are enough mergers to account for the present abundance of
spheroids or whether all gas-rich or elliptical-producing mergers
undergo an infrared luminous starburst phase.  The galaxy merger rate
and its evolution has been estimated
\citep[e.g.,][]{XS91,Burkley94,Carlberg94,SR94,KW95,
Soares95,YE95,Patton97,ZK98,Toledo99,LeFevre00,
Patton00,Patton02,Conselice03,Xu04,Lin04,Bundy04,Conselice05,Bell06}, but the
observational uncertainty is large.  Moreover, the few observational
studies of the distribution of quasar host luminosities
\citep{Bahcall97,McLure99,Falomo01,Hamilton02,
Jahnke03,Dunlop03,Floyd04,VandenBerk05} have not yet determined
if this distribution is expected (as a subset of the merging
galaxy population), or whether it requires modes of quasar
fueling aside from mergers of systems having reservoirs of cold gas.
Some observations \citep{Straughn05} and theoretical analyses
\citep{H05a} suggest that the stages most likely to be identified as
mergers can be quite distinct from the merger-driven quasar phase,
further complicating such comparisons.

Indirect evidence that mergers play a role in transforming galaxies
also comes from studies of the stellar mass budget contained in
morphologically or color-selected sub-populations as a function of
redshift.  For example, \citet{Borch06} find that the stellar mass in
red galaxies (known to be spheroid dominated) grows with time from
$z\sim 1$ to the present.  Furthermore, measurements of galaxy stellar
mass functions separated by color or by morphology
\citep{Bundy05a,Bundy05b,Franceschini06,Pannella06} show that the
``transition mass,'' above which red galaxies dominate the mass
function, increases with redshift, following a general trend of
``cosmic downsizing.'' As low-mass, red galaxies build up, the mass of
the largest star-forming galaxies, although less well-constrained,
decreases correspondingly, further supporting the view that star
formation is ``quenched'' in these systems and that they move onto the
red sequence \citep{Bundy05b}.  By comparing estimates of the
transition or quenching mass to the characteristic masses of quasars
(corresponding to the break in the observed quasar luminosity
function) and merging galaxies, \citet{H06} have shown that these
trace the same mass and evolve together over the range $0<z\lesssim3$,
implying that these objects are linked.  This connection motivates a
detailed analysis to understand the relationship between the merger
and quasar populations.

The contribution of mergers to the star formation density of the
Universe is not well known, especially at high redshifts
\citep[e.g.,][]{Brinchmann98,Menanteau01,Menanteau05,Bell05}.
Traditional observational measurements which require morphological
identification of star-forming galaxies are difficult to extend to
these redshifts and are likely incomplete \citep{Wolf05}, making
particularly valuable any independent constraints that do not require
direct morphological information, such as those from quasar
populations.  Even fewer estimates exist for the distribution of star
formation rates in merging systems \citep[e.g.,][]{Bell05}, and
whether or not the measurements are consistent with gas-rich mergers
passing through a starburst phase and building up the stellar mass
observed in the elliptical galaxy population.

The significant uncertainties in the relation of elliptical galaxy,
merger remnant, LIRG/ULIRG, merging (peculiar/interacting) galaxy,
starburst, and quasar populations are related to the same underlying
issue: Are the luminosity and mass distributions of quasars, quasar
host galaxies, merging galaxies, starbursts, and ellipticals
self-consistent?  If the origin of red/elliptical galaxies primarily
involves gas-rich mergers, with an intermediate starburst phase which
is terminated by feedback from black hole growth dispersing gas, then
these populations must track each other at all redshifts.

A long-standing obstacle to relating these populations has been the
lack of theoretical models which incorporate the relevant physical
processes.  Large-scale cosmological simulations
\citep[e.g.,][]{Cen94,Zhang95,Hernquist96,Katz96a,Katz96b,Navarro96,
Croft98,Croft99,Croft02,Dave99,McDonald00,McDonald04,Hui01,Viel03,Viel04}
and semi-analytic models
\citep[e.g.,][]{KH00,Cole00,Somerville01,V03,WL03,Granato04,Somerville04a,Baugh05,Croton05}
have provided a framework for understanding the intergalactic medium
and large-scale structure formation, but either do not resolve or do
not follow the detailed effects of star formation, supernova feedback,
black hole accretion and feedback, and dust obscuration, which are
critical for inferring the relation between e.g.\ quasar and starburst
populations and merger rates.  Instead, these processes are described
by idealized, tunable parameterizations rather than being modeled
dynamically.  Ideally, the prescriptions should be determined from
physically motivated models, which have been tested against
observations.

Recently, progress has been made in this direction by using
high-resolution simulations of individual galaxies undergoing mergers
that incorporate the effects of star formation and pressurization of
multi-phase interstellar gas by supernova feedback \citep{SH03} and
feedback from black hole growth \citep{SDH05b}.  In this manner, Di
Matteo et al. (2005) and Springel et al. (2005a,b) have shown that the
gas inflows produced by gravitational torques during a merger
(e.g. Barnes \& Hernquist 1991, 1996) both trigger starbursts
(e.g. Mihos \& Hernquist 1994a, 1996) and fuel rapid black hole
growth.  During most of this phase, the black hole is heavily obscured
\citep{H05a}, but becomes briefly visible as a bright optical quasar
when feedback expels the surrounding gas.  As the gas is heated and
dispersed, the accretion rate declines, leaving a dead quasar in an
ordinary galaxy.  The self-regulated nature of this process explains
observed correlations between black hole mass and properties of normal
galaxies \citep{DSH05}, as well as the color distribution of
ellipticals \citep{SDH05a}.

Previously (Hopkins et al.\ 2005a-d, 2006a-c; Lidz et al.\ 2006), we
have shown that this time evolution yields a self-consistent model for
many observed properties of quasars, with no dependence on a
particular prior cosmological distribution.  In particular, the
simulations imply a new description of quasar lifetimes where the
lifetime of a given source depends on both the instantaneous {\it and}
peak luminosities, so that for a given peak luminosity the lifetime
{\em increases} at lower luminosities -- i.e.\ quasars spend most of
their lives in phases fainter than their peak luminosities.  This
leads to a different interpretation of the quasar luminosity function
than is implied by idealized models of quasar lightcurves (in which
quasars grow/decay in a step function or pure exponential manner);
namely that the steep, bright end of the quasar luminosity function
traces quasars accreting at high Eddington ratios near their peak
luminosities, but the shallow, faint end is dominated by quasars with
high peak luminosities but which are seen in less luminous states.
The distribution of quasar birthrates as a function of their peak
luminosities (final black hole masses), which is directly related to
e.g.\ the gas-rich galaxy merger rate as a function of final galaxy
mass, is {\em peaked} at a luminosity/mass corresponding to the break
in the observed quasar luminosity function.  Objects near the peak in
this distribution dominate the observed faint-end quasar luminosity
function in their fainter, below peak phases.

In \citet{H05f}, we used our modeling to predict the distribution of
remnant elliptical galaxies formed in mergers.  Because the spheroid
stellar mass or velocity dispersion correlates with the final black
hole mass in our simulations, we can use our model of quasar lifetimes
and our determination of the birthrate of quasars as a function of
their peak luminosity (final black hole mass) to infer the rate at
which spheroids with given properties are formed in mergers as a
function of e.g.\ mass, velocity dispersion, size, and redshift.  In
this manner, we reproduce spheroid mass, luminosity, and velocity
dispersion functions, age distributions, mass-radius and
luminosity-size relations, mass-to-light ratios, color-magnitude
relations, and distributions of young (blue) spheroids at redshifts
$z=0-6$.  The co-evolution of star formation, black hole growth, and
obscuration is a key element in this analysis, and corresponding
predictions made either neglecting the role of black hole feedback in
terminating star formation or by modeling the quasar lightcurve in an
idealized manner are inaccurate.

In principle, the merger hypothesis provides a 
framework for describing the co-formation of quasars and spheroids
that reproduces a wide range of observations.  To investigate this
further, here we examine the statistics and properties of
merging systems.  In particular, we use our earlier modeling to study
the implied relation between the quasar luminosity function (QLF),
merging galaxy luminosity function (MGLF), quasar host galaxy
luminosity function (HGLF), distribution of star formation rates in
mergers (star formation rate function; SFRF), and the cumulative
merger rates and star formation rate density triggered by mergers.

In \S~\ref{sec:sims}, we describe our simulations, and in
\S~\ref{sec:guts} we use them to determine the evolution of near-IR
and optical/UV luminosities (\S~\ref{sec:guts.lum}), colors
(\S~\ref{sec:guts.colors}), and star formation rates
(\S~\ref{sec:guts.sfr}) during mergers, as a function of host galaxy
properties.  In \S~\ref{sec:methods} we describe our methodology for
mapping between various distributions
(\S~\ref{sec:comparisons.method}), and discuss the impact and
importance of possible systematic effects
(\S~\ref{sec:comparisons.systematics}).  In \S~\ref{sec:comparisons}
we use our modeling to predict the quasar luminosity function from the
observed merging galaxy function (at the same redshift), and vice
versa (\S~\ref{sec:comparisons.qlf.mglf}), in both $K$-band and the
optical/UV.  We use this to predict the underlying birthrate of
spheroids/quasars with a given final mass/peak luminosity, and compare
this with the observed merger mass functions
(\S~\ref{sec:comparisons.mf.lf}).  Likewise, we use this to predict the
distribution of star formation rates in mergers, independently from
the observed quasar and merger luminosity functions, and compare with
the observed star formation rate functions
(\S~\ref{sec:comparisons.sfr}).  In \S~\ref{sec:evolution}, we use the
observed evolution in the quasar luminosity function to predict the
evolution of the merging galaxy luminosity function and star formation
rate function, as well as integrated quantities such as the
luminosity, mass, and star formation rate densities in mergers, at
redshifts $z=0-6$.  In \S~\ref{sec:HGLF} we consider the relation
between the MGLF and HGLF, predicting the quasar host galaxy
luminosity function, the joint distribution of observed quasar and
quasar host galaxy luminosities, and their co-evolution. Finally, in
\S~\ref{sec:discuss}, we summarize our results and discuss their
implications for the merger hypothesis, various theoretical models,
and future observations.

Our approach is described in detail in
\S~\ref{sec:sims}-\ref{sec:methods}, but readers interested in the
main results may wish to skip to \S~\ref{sec:comparisons} or
\S~\ref{sec:HGLF}, where we compare and test the relation between
various stages and measures of merger-driven activity, and
\S~\ref{sec:evolution}, which uses these relations to predict
evolution of these measures with redshift.

Throughout, we adopt a $\Omega_{\rm M}=0.3$, $\Omega_{\Lambda}=0.7$,
$H_{0}=70\,{\rm km\,s^{-1}\,Mpc^{-1}}$ cosmology. All magnitudes are
Vega magnitudes, unless otherwise stated.

\section{The Simulations}
\label{sec:sims}

Our merger simulations were performed with the parallel TreeSPH code
{\small GADGET-2} \citep{Springel2005}, based on a fully conservative
formulation \citep{SH02} of smoothed particle hydrodynamics (SPH),
which conserves energy and entropy simultaneously even when smoothing
lengths evolve adaptively (see e.g., Hernquist 1993b, O'Shea et
al. 2005).  Our simulations account for radiative cooling, heating by
a UV background (as in Katz et al. 1996, Dav\'e et al. 1999), and
incorporate a sub-resolution model of a multiphase interstellar medium
(ISM) to describe star formation and supernova feedback \citep{SH03}.
Feedback from supernovae is captured in this sub-resolution model
through an effective equation of state for star-forming gas, enabling
us to stably evolve disks with arbitrary gas fractions (see,
e.g. Springel et al.\ 2005b; Springel \& Hernquist 2005; Robertson et
al. 2004, 2005).  This is described by the parameter $\qeos$,
which ranges from $\qeos=0$ for an isothermal gas with effective
temperature of $10^4$ K, to $\qeos=1$ for our full multiphase model
with an effective temperature $\sim10^5$ K.

Supermassive black holes are represented by ``sink'' particles
that accrete gas at a rate $\Mdot$ estimated from the local gas
density and sound speed using an Eddington-limited prescription based
on Bondi-Hoyle-Lyttleton accretion theory.  The bolometric luminosity
of the black hole is taken to be $\Lbol=\epsilon_{r}\dot{M}\,c^{2}$,
where $\epsilon_r=0.1$ is the radiative efficiency.  We assume that a
small fraction (typically $\approx 5\%$) of $\Lbol$ couples dynamically
to the surrounding gas, and that this feedback is injected into the
gas as thermal energy, weighted by the SPH smoothing kernel.  This
fraction is a free parameter, which we determine as in \citet{DSH05}
by matching the observed $M_{\rm BH}-\sigma$ relation.  For now, we do
not resolve the small-scale dynamics of the gas in the immediate
vicinity of the black hole, but assume that the time-averaged
accretion rate can be estimated from the gas properties on the scale
of our spatial resolution (roughly $\approx 20$\,pc, in the best
cases).

The progenitor galaxy models are described in
\citet{SDH05b}, and we review their properties here.  For each
simulation, we generate two stable, isolated disk galaxies, each with
an extended dark matter halo with a \citet{Hernquist90} profile,
motivated by cosmological simulations (e.g. Navarro et al. 1996; Busha
et al. 2005), an exponential disk of gas and stars, and (optionally) a
bulge.  The galaxies have total masses $M_{\rm vir}=V_{\rm
vir}^{3}/(10GH_{0})$ for $z=0$, with the baryonic disk having a mass
fraction $m_{\rm d}=0.041$, the bulge (when present) having $m_{\rm
b}=0.0136$, and the rest of the mass in dark matter.  The dark matter
halos are assigned a
concentration parameter scaled as in \citet{Robertson05b} appropriately for the 
galaxy mass and redshift following \citet{Bullock01}. 
The disk scale-length is computed
based on an assumed spin parameter $\lambda=0.033$, chosen to be near
the mode in the $\lambda$ distribution measured in simulations \citep{Vitvitska02},
and the scale-length of the bulge is set to $0.2$ times this.

Typically, each galaxy initially consists of 168000 dark matter halo
particles, 8000 bulge particles (when present), 40000 gas and 40000
stellar disk particles, and one BH particle.  We vary the numerical
resolution, with many simulations using twice, and a subset up to 128
times, as many particles. We choose the initial seed
mass of the black hole either in accord with the observed $M_{\rm
BH}$-$\sigma$ relation or to be sufficiently small that its presence
will not have an immediate dynamical effect, but we have varied the seed
mass to identify any systematic dependencies.  Given the particle
numbers employed, the dark matter, gas, and star particles are all of
roughly equal mass, and central cusps in the dark matter and bulge
are reasonably well resolved (see Figure 2 in Springel et
al. 2005b).

Our fitted quasar lifetimes and galaxy scaling relations
are derived from a series of several hundred simulations of colliding
galaxies, described in detail in \citet{Robertson05c,Robertson05b} and
\citet{H05e}.  We vary the numerical resolution, the orbit of the
encounter (disk inclinations, pericenter separation), the masses and
structural properties of the merging galaxies, initial gas fractions,
halo concentrations, the parameters describing star formation and
feedback from supernovae and black hole growth, and initial black hole
masses.

The progenitor galaxies have virial velocities $\vvir=80, 113, 160,
226, 320,$ and $500\,{\rm km\,s^{-1}}$, and redshifts $z=0, 2, 3, {\rm
and}\ 6$, and our simulations span a range in final black hole mass
$\mbh\sim10^{5}-10^{10}\,M_{\sun}$.  The extensive range of conditions
probed provides a large dynamic range, with final spheroid masses
spanning $\mbul\sim10^{8}-10^{13}\,M_{\sun}$, covering essentially the
entire range of the observations we consider at all redshifts, and
allows us to identify any systematic dependencies in our models.  We
consider initial disk gas fractions (by mass) of $\fgas = 0.2,\ 0.4,\
0.8,\ {\rm and}\ 1.0$ for several choices of virial velocities,
redshifts, and ISM equations of state.  The results described in this
paper are based primarily on simulations of equal-mass mergers;
however, by examining a small set of simulations of unequal mass
mergers, we find that the behavior does not change dramatically for
mass ratios down to about 1:3 or 1:4. This range is appropriate to the
observations of merging galaxies used in this paper, which are
restricted to `major' merger events.

\section{Luminosities, Colors, and Star Formation Rates of Mergers}
\label{sec:guts}

\subsection{Host Galaxy Luminosities}
\label{sec:guts.lum}

To de-convolve the observed pair or merging galaxy luminosity function
(MGLF) and infer a merger rate as a function of galaxy stellar mass,
we must first describe the starburst lightcurves.  In the simulations,
star formation is tracked self-consistently, with stellar ages and
stellar metallicities determined directly from the star-forming gas,
which itself is enriched by star formation.  We generally assume that
pre-existing stars (i.e.\ initial bulge or disk stars when the gas
fraction is less than unity) were formed with a constant star
formation rate prior to the starting point of the simulation, with
metal enrichment calculated self-consistently from the initial
primordial gas. Considering instead exponentially declining star
formation rates (``$\tau$-models'') or late-type star formation
histories fitted from observations
\citep[e.g.,][]{Kauffmann03b,Hammer05} makes relatively little
difference, and in any case stars formed over the course of the
simulations usually dominate the observed luminosity.  From the
predicted ages and metallicities of the stars in our simulations, we
use the stellar population synthesis models of \citet{BC03} assuming a
\citet{Salpeter55} initial mass function, to measure the luminosity in
a given band.

We determine attenuation in the manner described in \citet{H05a,H05e}.
Briefly, we calculate the column density along $\sim1000$
lines-of-sight to each simulation stellar particle at each time, 
using the SPH formalism and multiphase ISM equation of
state of the simulations \citep{SH03} to determine the density,
metallicity, and ionization state of gas in the ``warm/hot'' ISM
through which the majority of sightlines will pass (i.e.\ neglecting
mass collapsed in cold clouds).  We adopt a gas-to-dust ratio scaled
by metallicity \citep[e.g.,][]{Bouchet85} and normalized to that of
the Milky Way, $A_{B}/\nhi = 8.47\,(Z/0.02)\times10^{-22}\,{\rm
cm^{2}}$, and a Small Magellanic Cloud-like reddening curve (as
suggested by observations for the host galaxies of quasars, e.g.\
Hopkins et al.\ 2004, Ellison et al.\ 2005, although the 280\,nm
cross-section is decreased by only $6\%$ if we instead assume a Milky
Way-like reddening curve) with the form from \citet{Pei92}, to
attenuate the intrinsic spectrum of each stellar particle.  We do not
perform a full radiative transfer calculation, and therefore do not
model scattering or re-processing of radiation by dust in the
infrared.  However, we compare the column densities to quasars
calculated by variations of this method in \citet{H05b} (see their
Figures 1, 5, \& 6) and find that typical uncertainties are a factor
$\sim2-3$ in $\nh$, generally smaller than the variation across
different simulations and viewing angles.  Furthermore, we find below
that our results for the visible (attenuated) luminosities agree well
with those expected from the calculations in \citet{Jonsson05} which
employ a complete Monte Carlo radiative transfer model.

This modeling effectively captures obscuration by gas and dust in the
``typical'' ISM, but does not, however, directly resolve individual
molecular clouds or e.g.\ dense nuclear concentrations of molecular
gas, which have important effects on the mid and far-IR spectra of the
nuclear starbursts of LIRGs and ULIRGs.
These effects are unimportant
in our analysis, for several reasons.  First, the duration of the
starburst is short ($\sim10^{8}\,$yr) compared to the total
time during which a merger will be identified ($\sim10^{9}\,$yr),
and the fraction of the total stellar mass formed therein is small.
Most of the time the merger will be a weakly-interacting pair or
post-merger remnant, so corrections to luminosity or mass functions 
will in any case be small,
$\lesssim10\%$.  Moreover, surveys identifying mergers by pair counts
or tidal features with which we compare generally do not include the
heavily obscured starburst stage.

Second, we explicitly confine ourselves to consider the near-IR
through near-UV, where the effects of these obscuration mechanisms are
minimized.  For typical ULIRGs, central luminosities can be heavily
extinguished with $A_{\lambda}\sim50$ even in $K$-band; but as a
consequence the integrated luminosity is dominated by stars in more
extended regions, which are obscured by the relatively quiescent ISM
with a much lower effective total or ``screen'' extinction of
$\sim0.7$ magnitudes \citep[e.g.,][]{Lonsdale06}, similar to that in
our simulations. Third, re-radiation is relatively unimportant in the
bands we consider, and at the point of the nuclear starburst, even the
typical ISM densities are sufficient to render the most dense regions
optically thick. At such a point, the obscuration mechanism and the
``degree'' of optical thickness are unimportant.  Finally, we have
compared with full radiative transfer models in the far-IR through UV,
including dust heating and considering different sub-resolution
models for the detailed clumping of molecular gas (Narayanan et
al.\ 2006; Chakrabarti et al. 2006), and find little
change in the total luminosity in the bands we consider here,
even during peak starburst stages.

\subsubsection{K-band Luminosities}
\label{sec:guts.lum.K}
%\clearpage
\begin{figure}
    \scalesinglefig
    \centering
    \plotone{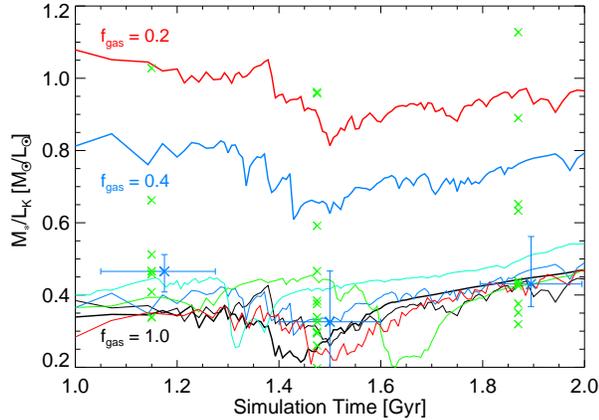}
    \caption{Ratio of total stellar mass to total $K$-band luminosity
    as a function of time in three representative simulations with gas
    fractions $\fgas=0.2,\ 0.4,\ 1.0$ as labeled (thick lines).  For
    $\fgas=1.0$, several simulations with $\vvir=80-320\,{\rm
    km\,s^{-1}}$ and $\qeos=0.25-1$ are shown (thin lines). Green 
    points show observed ULIRGs (expected to correspond to high-$\fgas$ mergers), 
    with blue points showing the median and interquartile range, from the 
    observed merging ULIRG pair sample of \citet{Dasyra06} ($t\sim1.2$\,Gyr; horizontal 
    error roughly illustrates the times corresponding to observed separations), 
    the late-stage ULIRG sample of \citet{Tacconi02} ($t\sim1.5\,$Gyr), and 
    the merger-remnant LIRG/ULIRG sample of \citet{RJ04} ($t\sim1.9\,$Gyr). 
    The $K$-band 
    mass-to-light ratio depends weakly on gas fraction in a 
    systematic manner, but not on other varied merger properties, 
    and agrees well with typical observed values.
    \label{fig:K.ML}}
\end{figure}
%\clearpage

We first consider the $K$-band galaxy luminosity in mergers.  In each
simulation, we calculate and plot the total stellar mass divided by
total $K$-band luminosity as a function of time.
Figure~\ref{fig:K.ML} shows the result for several representative
cases, some with identical virial velocities, but a varying gas
fraction, and others with the same $\fgas$, but either different
$\vvir$ or $\qeos$.  There is no clear systematic difference in $\mlk$
at fixed gas fraction (we have also varied e.g.\ initial black hole
mass, presence or absence of initial bulges, and orbital parameters
and find the same).  However, there is a systematic dependence on the
gas fraction, so that $\mlk$ is somewhat lower for high-$\fgas$
systems which are dominated by younger stellar populations.

We can estimate this effect quantitatively as a function of $\fgas$
with a model in which a galaxy of total final stellar mass
$\mtot$ forms a fraction $\fgas$ of its mass in the merger (i.e.\
assuming the limit of efficient conversion of gas to stars) with
mass-to-light ratio $(M/L_{\rm K})_{\rm new}$, and the remaining mass
coming from an older population with $(M/L_{\rm K})_{\rm old}$.  This
gives
\begin{equation}
\mlkfrac = \frac{(M/L_{\rm K})_{\rm new}}{\fgas + (1-\fgas)\,
\frac{(M/L_{\rm K})_{\rm new}}{(M/L_{\rm K})_{\rm old}}}.
\label{eqn:K.ML}
\end{equation}
Fitting to the mean $\mlk$ in our mergers as a function of $\fgas$
gives $(M/L_{\rm K})_{\rm new}\approx0.4$ and $(M/L_{\rm K})_{\rm
old}\approx1.4$, corresponding to stellar populations with mean ages
$\sim0.5$ and $5$\,Gyr, respectively, reasonable values for observed
mergers.  This equation gives the mean $\mlk$ accurate to
$\sim10-20\%$; i.e.\ comparable to the scatter in $\mlk$ in a given
merger and across mergers with varied parameters (but similar gas
fractions). Note that the dependence is fairly weak, as a factor of 2
change in gas fraction (e.g.\ $\fgas=0.1\rightarrow0.2$ or
$\fgas=0.2\rightarrow0.4$) results in only a $\sim20\%$ change in
$\mlk$, generally smaller than the relevant observational
uncertainties.

The $K$-band total mass-to-light ratio is also relatively constant
with time in a merger.  In part, this owes to the weak dependence of
$\mlk$ on stellar population age and the continuous star formation
throughout the merger.  For example, given a constant star formation
rate (SFR) beginning at $t=0$, and ignoring older stars (i.e.\ roughly
the $\fgas=1$ case), we expect $\mlk$ to increase by only
$\sim40\%$ from $t=1\,$Gyr to $t=2\,$Gyr, as in Figure~\ref{fig:K.ML}.
Moreover, as seen in simulations \citep{H05e,Jonsson05} and
observations of local ULIRGs such as Arp 200 \citep{Doyon94,Lonsdale06},
increased densities in the merger give
rise to both elevated star formation but also enhanced obscuration,
resulting in a flat observed host galaxy luminosity during these
times.

Unfortunately, without a large sample of mergers or a reliable means
to determine the exact merger stage or pre-merger $\fgas$, comparing
our predicted $\mlk$ as a function of time with observations is not
possible, but we can check for consistency.  In Figure~\ref{fig:K.ML},
we plot $\mlk$ of observed merging ULIRG pairs with masses determined
from dynamical measurements in \citet{Dasyra06} at $t\sim1.2$\,Gyr,
corresponding roughly to the stage at which the merging nuclear pair
separations are comparable.  Green crosses show individual objects,
blue points the median and interquartile range, with horizontal errors
heuristically corresponding to times in which a given merger would
appear in the observed sample.  At $t\sim1.5\,$Gyr we show the same,
for the late-stage ULIRG sample of \citet{Tacconi02}, and at
$t\sim1.9\,$Gyr, for the post-merger LIRG/ULIRG sub-sample of
\citet{RJ04}.  At each stage, the typical values and scatter in $\mlk$
correspond well with that expected, especially given the additional
scatter in observed samples from measurement errors and a range in
observed $\fgas$ (note that the observed samples are all selected to
be ULIRGs, so we expect they will correspond to the most gas-rich
mergers we simulate). The later-stage (non-LIRG) merger remnant sample
of \citet{RJ04} forms a continuum in $\mlk$ from these values to those
typical of old ellipticals $\mlk\sim1.4$, presumably as these
populations age and redden.

\subsubsection{280\,nm Luminosities}
\label{sec:guts.lum.280}

\begin{figure}
    \scalesinglefig
    \centering
    \plotone{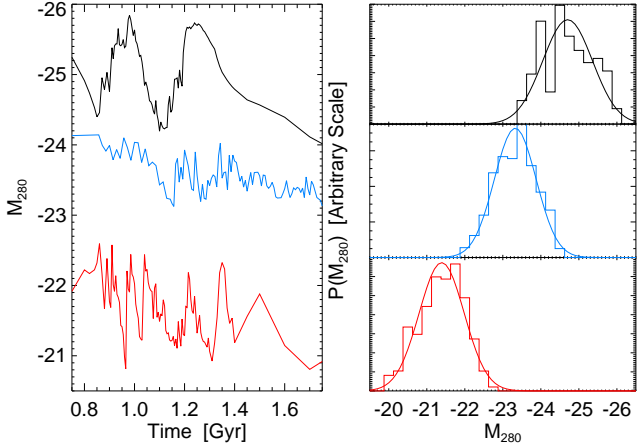}
    \caption{Left: Median observed (attenuated) magnitude at 280\,nm
    as a function of time during the peak merger stages from three
    representative simulations with $\vvir=80,\ 160,\ {\rm and}\
    320\,{\rm km\,s^{-1}}$ (red, blue, and black, respectively).
    Right: Distribution of observed magnitudes (each weighted by
    the amount of time across all sightlines that the galaxy is
    observed at a given $\muv$) for each simulation (histograms of
    corresponding colors), with best-fit Gaussians (smooth curves).
    The distribution of optical/UV luminosities is relatively constant throughout the 
    merger, with enhanced star formation and obscuration offsetting one another, 
    and is well-fitted by a Gaussian PDF. 
    \label{fig:uv.ex}}
\end{figure}

Next, we examine a synthetic UV band centered at $280$\,nm, with a
40\,nm width, for comparison with the results of \citet{Wolf05} from
the GEMS and GOODS surveys.  This band is normalized to the solar
luminosity in the 280/40 passband, which is $M_{\sun,\, 280}=6.66$ in
Vega units or $L_{\sun,\, 280}=2.56\times10^{10}\,{\rm W/Hz}$.  Here,
we use the Starburst99 population synthesis models \citep{starburst99}
to calculate the UV spectra of the stellar particles.
Figure~\ref{fig:uv.ex} shows the outcome of this calculation for three
representative simulations, with $\fgas=1.0$, $\zgal=0$, $\qeos=1.0$,
and varying $\vvir$, as labeled.  The median observed $\muv$ as a
function of time during the merger is shown for each simulation (left
panel).  Again, the effects of star formation and dust obscuration
tend to offset one another, and the host galaxy luminosity varies
within a factor $\lesssim2$, i.e.\ within 1 magnitude, throughout the
merger, even though the {\em intrinsic} (un-attenuated) UV luminosity
can rise by a much larger factor.

%\clearpage
%\clearpage

We calculate (Figure~\ref{fig:uv.ex}, right panel) the probability of
observing a given $\muv$, i.e.\ the total time a given $\muv$ is
visible integrated over all sightlines and over the duration of the
merger.  The results are shown as histograms of the corresponding
color for each merger in the left panel (with the best-fit Gaussian).
In each case, the PDFs have a well-defined peak and narrow
($\sim0.6\,$mag) width in $\muv$, emphasizing that the light curves
are quite flat. The Gaussian fits allow us to statistically
characterize the lightcurves of our simulations with a mean $\muv$ and
rms dispersion about this mean, $\suv$.

%\clearpage
\begin{figure}
    \scalesinglefig
    \centering
    \plotone{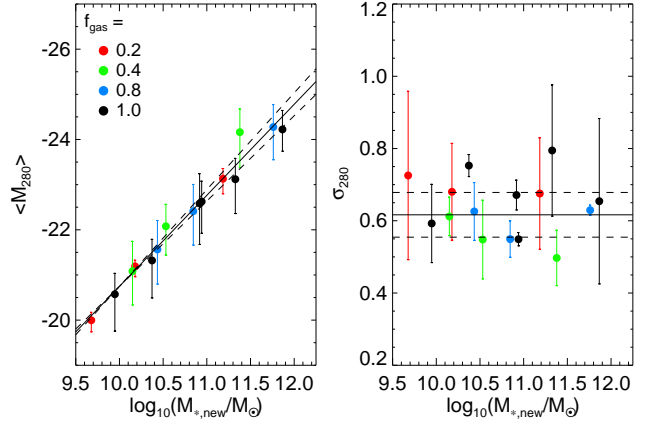}
    \caption{Left: Mean observed 280\,nm magnitude over the course of
   a merger for simulations with $\vvir=80-320\,{\rm
   km\,s^{-1}}$ and gas fractions as labeled (circles), as a function
   of the mass of stars formed during the merger,
   $\mnew$. Vertical errors show the range allowing for a factor of 3
   uncertainty in the calculated column density.  Solid line shows the
   best-fit power law, dashed lines the $1\sigma$ range of this fit.
   Right: Dispersion in 280\,nm magnitude over the course of a merger,
   for the same simulations (symbols as in left panel). Solid line shows
   the best-fit constant $\suv$ and dashed lines the $1\sigma$ range.
   The PDF for observing a merger at a given luminosity has a constant shape and width, 
   with the mean optical/UV luminosity scaling with the mass of stars forming 
   in the merger.
    \label{fig:uv.scaling}}
\end{figure}
%\clearpage

In Figure~\ref{fig:uv.scaling}, we quantify the dependence of the mean
observed 280\,nm luminosity on host galaxy properties.  We include a
subset of our simulations with various values of $\fgas$, as labeled,
and for each, $\vvir=113,\
160,\ {\rm and}\ 320\,{\rm km\,s^{-1}}$ and $\qeos=1.0$.  For
$\fgas=1.0$, we also calculate the results for $\vvir=80,\ 226\ {\rm
km\,s^{-1}}$, $\qeos=1.0$ and for $\vvir=160\,{\rm km\,s^{-1}}$,
$\qeos=0.25$, to demonstrate that these variations do not alter our
conclusions.  We plot the mean $\muv$ and rms dispersion $\suv$ for
each simulation. The vertical error bars show the range
resulting from a systematic increase or decrease in our
estimated column densities (roughly parameterizing the maximal
uncertainty for our calculation of $\nh$ from Hopkins et al.\
2005b). We expect the UV luminosity will be dominated by young
stars, and therefore show the mean $\muv$ as a function of the mass of
stars formed during the merger, $\mnew$.  Unsurprisingly, the two are
well correlated, and the scaling does not depend on gas fraction,
virial velocity, or ISM gas equation of state.  Fitting a relation of
the form
\begin{equation}
\muv=M_{280,\, 0}+\alpha\,\log_{10}{\Bigl(}\frac{\mnew}{10^{10}\,M_{\sun}}{\Bigr)}
\label{eqn:M280.fit}
\end{equation}
gives a best fit $M_{280,\, 0}=-20.7\pm0.1$, $\alpha=-2.01\pm0.13$,
shown as a solid black line.
In terms of $\luv$,
\begin{equation}
\luv=L_{280,\, 0}\,{\Bigl(}\frac{\mnew}{10^{10}\,M_{\sun}}{\Bigr)}^{0.80\pm0.05}, 
\label{eqn:L280.fit}
\end{equation}
where $L_{280,\, 0} = 5.06\pm0.45\times10^{10}\,L_{\sun,\ 280}$.
The dispersion $\suv$ has no
significant dependence on stellar or black hole mass, gas fraction, or other 
host properties. The solid line (dashed lines)
shows the best fit ($\pm1\sigma$) mean $\suv=0.62\pm0.06$ ($0.25\,$dex 
in $L_{280}$).  There are limited observations to compare against,
but these values are consistent with the colors and inferred 
star formation properties in e.g.\ \citet{Bell05b}. 

The scaling of $\muv$ with host properties can be understood in terms
of a model assuming efficient conversion of gas to stars.  Given that
general trends in star formation rate as a function of time are
qualitatively similar regardless of the total number of stars formed,
the intrinsic (un-attenuated) UV luminosity (henceforth $L_{\rm int}$)
should scale proportionally with the mass of new stars formed, $L_{\rm
int}=\mluveff^{-1} \mnew$, where $\mluveff$ is an effective
mass-to-light ratio depending on the star formation history.  The
timescale of $\sim 1$ Gyr during which the galaxies will be visible as a
merger, or alternatively the fit to $\mlk$ from our simulations or
observations \citep{Tacconi02}, imply
$\mluveff\approx0.08\,M_{\sun}/L_{\sun,\ 280}$, corresponding to a
population with an age of $0.5\,$Gyr.

However, the observed luminosity does not scale as steeply as $L_{\rm
int}\propto\mnew$, owing to higher attenuation as the gas mass
increases. \citet{Jonsson05} consider lightcurves of merging galaxies
using a smaller set of simulations (without black holes), but
with a full Monte Carlo radiative transfer code, and demonstrate that
it is reasonable to assume that the newly-formed stars and
star-forming gas (which produces most of the obscuration) are
well-mixed, giving an observed luminosity $\luv = L_{\rm
int}\,\tau^{-1}\,(1-e^{-\tau})$ \citep[e.g.,][]{Calzetti94}, where
$\tau$ is an effective mean optical depth.  From their fits to the
attenuation in bolometric, SDSS $u$-band, and GALEX NUV luminosities
we estimate $\tau\approx \tau_{0}\,(M_{g}/10^{11}\,M_{\sun})^{0.16}$,
where $\tau_{0}\approx2-3$ and $M_{g}\approx \mnew$ is the gas mass,
which they show can be roughly understood in terms of how both the SFR and
obscuration scale with local gas density.  Given $\tau\gtrsim1$, the
attenuation goes as $\luv\propto \tau^{-1}$, and we can combine the
scalings to obtain the expected relation
\begin{equation}
\luv\approx 5.8-8.7\times10^{10}\,L_{\sun,\ 280}
\,{\Bigl(}\frac{\mnew}{10^{10}\,M_{\sun}}{\Bigr)}^{0.84}, 
\label{eqn:L280.theory}
\end{equation}
in good agreement with the scaling of $\luv$ measured
in the simulations. That these scalings agree emphasizes that
contributions to the observed $\luv$ from older stellar populations
are relatively small ($\lesssim30\%$).  Furthermore, this suggests
that our results using just the calculated column densities (i.e.\
ignoring scattering processes) would not be significantly changed by
a more sophisticated treatment of radiative transfer.

Finally, the conversion of gas to stars in mergers is efficient, i.e.\
$\mnew\approx\fgas\mtot=\fgas M_{\ast}$ (true to $\lesssim10\%$ over
$\fgas\sim0.2-1.0$), where for simplicity we subsequently denote the
final, total stellar spheroid mass as $M_{\ast}=\mtot$.  Our
calculations then describe the statistics of $K$-band and 280\,nm
luminosities in mergers as a function of $\fgas$ and $M_{\ast}$, with
no systematic dependence (quantified in this manner) on the other
parameters we have varied.

\subsubsection{Other Optical and Near-IR Luminosities}
\label{sec:guts.lum.other}

For future reference, we perform the above set of calculations
in several different wavebands (280\,nm, U, B, V, R, I, J,
H, K, and SDSS u, g, r, i, z), calculating for each the equivalent of
Equation~(\ref{eqn:M280.fit}), i.e.\ the best-fit relation between the
median magnitude in the given band ($M_{\rm BAND}$) and total stellar
mass formed in the merger ($\mnew$), as well as the dispersion in a
given merger about the expected median magnitude ($\sigma_{\rm
BAND}$). The results are shown in Table~\ref{tbl:magslookup}. We also
consider fits to $M_{\rm BAND}$ as a function of final black hole mass
$M_{\rm BH}$ or total final stellar mass $\mtot$ instead of
$\mnew$. For each, we show the resulting $\reducechi$, which
measures the goodness-of-fit.  Note that a small $\reducechi$
implies small scatter about the fitted relation, but does not imply
that there is no systematic dependence on certain external variables.

For example, in the optical/UV, the correlation between $M_{\rm BAND}$
and $\mnew$ or $\mtot$ is comparable, but as demonstrated above, there
is a systematic dependence on $\fgas$ if $M_{\rm BAND}$ is quantified
as a function of $\mtot$ (brighter optical/UV at higher gas
fractions). Note that $\sigma_{\rm BAND}$ also quantifies the degree
to which it is appropriate to assume that the luminosity in the given
band over the course of the merger is constant (a poor approximation
to the bolometric luminosity, but a surprisingly good one in the bands
we measure).  These should be of particular use in comparing
cosmological simulations and semi-analytic models with observed merger
populations, and enable future observations of merger luminosity
functions to easily estimate their contributions to remnant spheroid
and quasar populations.

\subsection{Colors and Color-Magnitude Relations}
\label{sec:guts.colors}
%\clearpage
\begin{figure}
    \scalesinglefig
    \centering
    \plotone{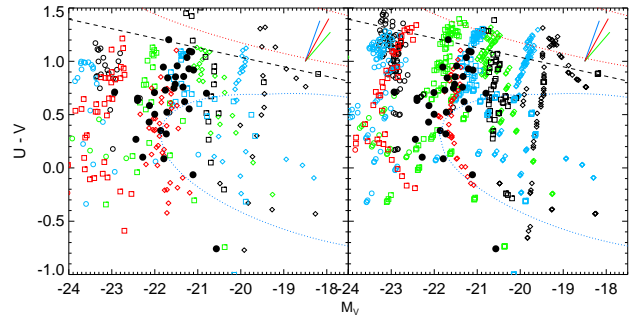}
    \caption{Color-magnitude relation at $\sim100$ random times and
    sightlines during each of several merger simulations with
    $\vvir=116,\ 160,\ 320\,{\rm km\,s^{-1}}$ (diamonds, squares, and
    open circles, respectively), $\fgas=0.2,\ 0.4,\ 0.8,\ 1.0$ (black,
    blue, green, and red, respectively), $\zgal=0$, $\qeos=1.0$. The
    left panel neglects dust attenuation, the right panel includes
    this effect. In each, solid lines show the impact of increasing
    metallicity (from $0.1\,Z_{\sun}-Z_{\sun}$; blue), age (from
    $0.5-1\,$Gyr; green), and column density (from
    $\nh=0-10^{21}\,{\rm cm^{-2}}$; red). Filled black circles show
    local ULIRGs from \citet{Surace00} and \citet{Arribas04},
    K-corrected following \citet{Bell05}. For comparison, dotted contours 
    (roughly) show the local red sequence (red) and blue cloud (blue), with the 
    ``valley'' separating the two from \citet{Bell05} (dashed). 
    The color-magnitude distribution of the simulations is similar to that 
    observed, with roughly equal contributions from stellar population 
    effects and variations in column densities.
    \label{fig:UV}}
\end{figure}
%\clearpage

In Figure~\ref{fig:UV}, we plot the $(U-V)$ vs.\ $M_{V}$
color-magnitude relation for each of several representative
simulations with different virial velocities and gas fractions
$\fgas=0.2-1.0$, fixing $\zgal$ and $\qeos$.  For each simulation, we
consider $\sim100$ randomly selected points in time at which the
merger is ``observed'' during the $\sim2\,$Gyr during which the system
shows evidence of merging, and at each point in time consider a random
sightline (uniformly sampling the unit sphere).  In the left panel, we
neglect the effects of dust attenuation.  We illustrate the
age/metallicity/obscuration degeneracy with the solid lines in the
upper right, which demonstrate how a point in the color-magnitude
space is moved (from the point of origin of the three lines) with
increasing mean stellar age, metallicity, or column density.

The wide scatter in $(U-V)$ colors in both panels demonstrates that
the effects of different mean stellar ages and metallicities
contribute strongly to the scatter in the color-magnitude
relation. The evolution is difficult to disentangle, as increasing age
and metallicity in later stages generally redden the merger, but the
final merger is associated with a starburst which briefly reduces the
mean (luminosity-weighted) age. Differential extinction actually
marginally decreases the color spread, as intrinsically blue periods
(i.e.\ times of peak star formation) tend to be heavily dust-reddened,
whereas intrinsically red periods (e.g.\ post-merger stages) are only
weakly affected.  Although the scatter in $(U-V)$ may derive largely
from stellar population effects, this does not imply that extinction
is unimportant. The mean effective total or ``screen'' reddening is
$\sim0.2-0.5$ magnitudes and during the peak starburst phase this
rises to $\gtrsim1-2\,$mag, similar to that observed
\citep[e.g.,][]{Genzel98}.  (Note, however, that the spatially resolved
extinction can be much greater in the very central regions during
these phases.)  Extinction also reddens objects out of the bright blue
($M_{V}\lesssim-21$, $(U-V)\lesssim0$) region of the color-magnitude
diagram (quantitatively, the fraction of points in this region drops
from $\sim10\%$ to $\sim1\%$ when attenuation is included).
% exact numbers $8.6^{+4}_{-2}\%$ and $1.3^{+1.2}_{-0.4}\%$, respectively
The faint, red region is not especially heavily populated, which is of
interest for optically or UV-selected samples, especially at high
redshift, which may not be sensitive to faint red mergers \citep[see
e.g.][for further discussion]{Bell05,Wolf05}. We emphasize though that
there is no real color-magnitude ``relation,'' and thus any physical
inferences from the observed color-magnitude distribution should be
considered with caution.

We plot for comparison (filled circles) the locations of local ULIRGs
from \citet{Surace00} and \citet{Arribas04} following
\citet{Bell05}. Quantitative comparison is difficult, but the observed
and simulated loci are similar, especially compared to the blue cloud
and red sequence on which the progenitor and remnant galaxies lie,
respectively \citep[see][]{SDH05a}.  We have also compared the
$(280-V)$, $(U-B)$, $(B-R)$, and $(R-K)$ colors with observed
color-magnitude distributions of the $z\approx0.7$ interacting sample
of \citet{Bell05} and \citet{Wolf05}, and the morphologically
irregular starburst sample of \citet{deMello05}, as well as the GEMS
host galaxy sample of active galactic nuclei (AGN) of
\citet{Sanchez04} (see e.g.\ Figure 7 of Bell et al.\ 2005), and find
similar broad agreement. The substantial scatter in metallicities
during mergers is also comparable to that in \citet{RJ05}, and rough
trends in obscuration and metallicity with stellar mass or luminosity
in the star-forming stages are similar to those observed
\citep{Burstein88,Worthey92,Faber95,Jorgensen97,Kuntschner00}.

\subsection{The Star Formation Rate Distribution in Individual Mergers}
\label{sec:guts.sfr}

Because the star formation rate is far from constant during a merger,
sharply peaking during the starburst phase and falling off
exponentially afterwards, with smaller peaks in earlier stages, a
given merger will not necessarily be observed near a single
characteristic SFR (as with e.g.\ the luminosities calculated above).
We can, however, consider the time a given simulation spends in a
given interval in SFR (i.e.\ the SFR-dependent ``lifetime'' of the
merger) and use this to quantify the distribution of SFRs in
mergers.  In detail, the time-behavior of the SFR, $\sfr$, is 
somewhat chaotic, with a systematic dependence on virial velocity, ISM
gas equation of state, galaxy gas fractions, and merging galaxy mass
ratio. However, when rescaled in terms of the {\em total} mass of new
stars formed during the merger, $\mnew$, the critical statistical
property of the SFR distribution, namely the time spent by a given
simulation in a given interval in $\sfr$, is essentially unaffected
by these systematics.

%\clearpage
\begin{figure}
    \scalesinglefig
    \centering
    \plotone{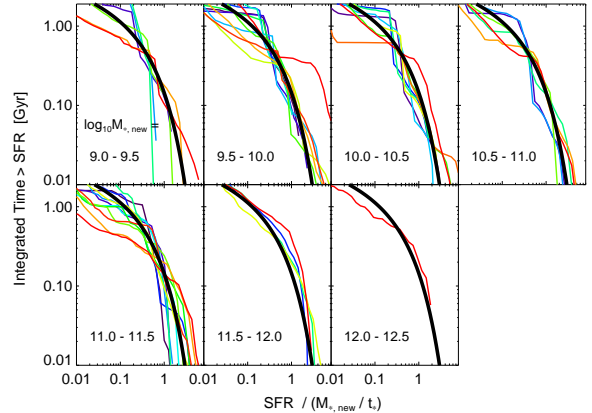}
    \caption{Integrated time above a given rescaled star formation rate 
    $\dot{M}/(\mnew/t_{\ast})$, where $\mnew$ 
    is the mass of new stars formed during the merger and $t_{\ast}=0.65\,$Gyr 
    is fitted. Results are shown for $\sim100$ simulations, in bins of 
    $M_{\ast,\,\rm new}$ as labeled. Different colors show simulations with 
    different values of $\vvir$, $\fgas$, $\qeos$, $\zgal$. The thick black line in 
    each interval shows the fitted analytical approximation of 
    Equation~(\ref{eqn:t.sfr}). 
    Quantified in this manner, the PDF for observing a given merger at some 
    instantaneous SFR is well-determined and independent of systematic effects from 
    varied quantities in the simulations. 
    \label{fig:t.sfr}}
\end{figure}
%\clearpage

In Figure~\ref{fig:t.sfr}, we determine the integrated time spent by
each of $\sim100$ simulations above a given SFR (from
$\sfr\sim10^{-4}-10^{3}\,{\rm M_{\sun}\,yr^{-1}}$), as a function of
that SFR, $\sfr$. We then rescale the SFR by dividing out a
characteristic SFR, $\mnew/\tsfr$, i.e.\ the total new stellar mass
formed during the merger divided by some characteristic starburst
timescale. For now, $\tsfr$ is arbitrary (we choose $\tsfr=0.65$\,Gyr,
for the reasons below).  We plot, for each simulation, the time spent
above a given dimensionless SFR ($\sfr/(\mnew/\tsfr)$). We show this
for the simulations in bins of $\log(\mnew)$ from
$\mnew\sim10^{9}-10^{13}$. In each bin, different colored lines
correspond to different initial conditions in the simulations, various
combinations of $\vvir=80-500\,{\rm km\,s^{-1}}$, $\qeos=0,\ 1$,
$\zgal=0,\ 2,\ 3,\ 6$, and $\fgas=0.2,\ 0.4,\ 0.8,\ 1.0$. Although
there is significant scatter in the time spent above some SFR between
different simulations, we find that once rescaled in this manner,
there is no systematic dependence on any of the varied quantities.

We fit the ``lifetime'' at a given SFR in Figure~\ref{fig:t.sfr} to an
analytical function following \citet{H05e},
\begin{equation}
\frac{{\rm d}t}{{\rm d}\log \sfr}=\tsfr\, \ln(10)\,
\exp{{\Bigl(}-\frac{\sfr}{\mnew / \tsfr}{\Bigr)}},
\label{eqn:t.sfr}
\end{equation}
where $\mnew$ is the total stellar mass formed in the merger, $\tsfr$
is a characteristic timescale, and the normalization $\tsfr\,\ln(10)$
is determined by the integral constraint
\begin{equation}
\int{\sfr\,{\rm d}t}=
\int{\sfr\,\frac{{\rm d}t}{{\rm d}\log \sfr}\,{\rm d}\log(\sfr)}=
\mnew .
\end{equation}
This gives a one-parameter function to describe the
statistical properties of the SFR distribution in mergers.

Fitting to the simulations, we find a best-fit
$\tsfr\approx0.65\pm0.10$\,Gyr, with no evidence for a strong
dependence of $\tsfr$ on $\mnew$ or other varied parameters. This is
comparable to the star formation timescale implied by observations of
starburst galaxies \citep{Kennicutt98} and mean luminosity-weighted
ages of merger stellar populations \citep{Tacconi02}.  We show the
prediction of this fit as the thick black lines in each panel of
Figure~\ref{fig:t.sfr}. The form of Equation~(\ref{eqn:t.sfr})
suggests an exponentially increasing/decreasing SFR into/out of the
starburst peak. However, too much detail should not be read into the
precise SFR as a function of time from these fits, as they both
average over the entire merger and do not capture a systematic
dependence such as the tendency of higher-mass systems to form a
larger fraction of their stars prior to the final merger
\citep{Robertson05c}.  As an aside, a similar functional form could be
used to describe quiescent SFRs in exponentially declining
``$\tau$-models,'' but in this case would be characterized by a much
lower typical SFR ($\sim \fgas M_{\rm disk} / 10\,{\rm Gyr}$) and
higher normalization $\sim\tau\ln{10}$ (where $\tau\sim4-8\,$Gyr is
the SFR timescale in stable disks; e.g.\ Li et al.\ 2005,\,2006).

\section{Methodology: Comparing Merger and Quasar Distributions}
\label{sec:methods}

\subsection{Converting Between Merger, Quasar, and SFR Distributions}
\label{sec:comparisons.method}

If the
merging galaxy luminosity function (MGLF) in a band $\nu$ is given by $\phi_{\rm
gal}(L_{\nu})\equiv{\rm d}\Phi/{\rm d}\log(L_{\nu})$, then
\begin{equation}
\phi_{\rm gal}(L_{\nu})=\int \frac{{\rm d}t}{{\rm d}\log(L_{\nu})}\,\dot{n}(M_{\ast})\,{\rm d}\log M_{\ast},
\label{eqn:phigal}
\end{equation}
where $\dot{n}(M_{\ast}) = {\rm d}n\,{\rm d}t^{-1}\,{\rm
d}\log(M_{\ast})^{-1}$ is the merger rate as a function of final
stellar mass, and ${\rm d}t/{\rm d}\log(L_{\nu})$ is the time the
merger is visible in a logarithmic interval in $L_{\nu}$.  Because, as
shown in \S~\ref{sec:guts.lum}, the merging galaxy luminosity is quite
flat in time, and the timescale over which merging galaxies are
identifiable as such ($\sim2$\,Gyr) does not depend strongly on the
galaxy masses over the range of interest, we can use ${\rm d}t/{\rm
d}\log(L_{\nu}) = P(L_{\nu}\,|\,M_{\ast})\,\tmerge$, where
$P(L_{\nu}\,|\,M_{\ast})$ is the probability (calculated in
\S~\ref{sec:guts.lum}) of being observed in a logarithmic interval in
$L_{\nu}$ over the merger timescale ($\tmerge$). The de-convolution of
the MGLF, $\phi(L_{\nu})$, with the PDF for observing a merger at a
given luminosity $P(L_{\nu}\,|\,M_{\ast})$ directly yields the merger
mass function for a given luminosity function.

Having considered the observed properties of mergers with some 
final spheroid mass $M_{\ast}$, we can use the scalings 
between black hole mass and host galaxy velocity dispersion
or stellar spheroid mass to relate each merger to an
associated black hole. 
The relation between final (post-merger) black hole mass and 
total stellar mass in our simulations is 
\begin{equation}
\mbh \approx 0.001\,M_{\ast},
\end{equation}
\citep{DSH05,Robertson05b}, with an approximately 
lognormal scatter of $\sim0.3\,$dex, essentially identical 
to observational estimates from e.g.\
\citet{MH03,HaringRix04} (accounting for the difference between 
virial and stellar mass). 
Knowing $\mbh(M_{\ast})$, we can convert
$\dot{n}(M_{\ast})$ to $\dot{n}(M_{\rm BH})$, 
the birthrate of quasars of a given relic mass in galaxy mergers, 
\begin{equation}
\dot{n}(\mbh)=\int P(\mbh\,|\,M_{\ast})\,\dot{n}(M_{\ast})\,{\rm d}\log M_{\ast} \, .
\end{equation}
We can then express the observed QLF in
terms of $\nLp$ and the quasar lifetime as a function of luminosity
(see Equation~[\ref{eqn:qso.lifetime}]), identical to the expression in
Equation~(\ref{eqn:phigal}) but with $M_{\ast}$ replaced by $M_{\rm BH}$.  

Given a model for the quasar lifetime ${\rm d}t/{\rm d}\log{(L_{\nu})}$, then, the QLF is 
determined. We generally consider two cases: first, 
a ``feast or famine'' or ``light-bulb'' model
for the quasar lifetime, following what has generally been adopted in
previous works \citep[e.g.,][]{SB92,KH00,HM00,WL03,V03,HQB04}.  In one
typical scenario of this type, the quasar turns on in a merger,
accretes at constant Eddington ratio $\dot{m}$, and then either
``turns off'' or exponentially decays in luminosity \citep{HL98}. 
The lightcurve is an exponential, $L\propto\exp(\pm t/t_{Q})$, 
and the lifetime as a function of luminosity is trivial, 
${\rm d}t/{\rm d}\log(L)=\ln10\ t_{Q}={\rm
constant}$ (for $L<L_{\rm peak}$), where $t_{Q}$ is a free parameter. 
Assuming an even simpler description in which
quasars ``turn on'' at fixed luminosity $L=L_{\rm peak}$ for a fixed
$t_{Q}$, produces nearly identical results. Generally, $t_{Q}$ is
either adopted from observations (with the loose constraint
$t_{Q}\sim10^{6}-10^{8}\,$yr; see Martini 2004 for a review) or
assumed to be the Salpeter (1964) time $t_{S}=4.2\times 10^{7}\,$yr 
($e$-folding time at $\dot{m}=1$ with 
radiative efficiency $\epsilon_{r}=0.1$), but we allow it to vary to provide
the best fit to the observed QLF.

In Hopkins et al.\ (2005a,b, 2006a,b), we compare these idealized scenarios to 
the light curves and lifetimes derived from our simulations and find that they 
provide a poor representation at any luminosity. 
The simulation lightcurves are complex, with multiple phases of activity,  
variability, and feedback-regulated decay, but their statistical 
nature can be described by simple forms \citep{H05b,H05e}.
The key feature of the quasar lifetime not captured by more idealized
models is that it {\em increases} with decreasing luminosity; i.e.\ a
given quasar spends more time at luminosities below its peak than at its
peak luminosity. In \citet{H05e,H05f} we use our simulations to 
calculate the differential quasar lifetime, i.e.\ the time spent in a given
logarithmic luminosity interval, and find it is well-fitted 
by a Schechter function,
\begin{equation}
\frac{{\rm d}t}{{\rm d}\log(L)}=t_{Q}\,{\Bigl(}\frac{L}{L^{\ast}_{Q}}{\Bigr)}^{-\alpha}\, \exp [-L/L^{\ast}_{Q}].
\label{eqn:qso.lifetime}
\end{equation}
Here, $L^{\ast}_{Q}$ is proportional to the {\em peak} quasar
luminosity $\Lp\sim L_{\rm Edd}(M_{\rm BH})$, $t_{Q}$ is a fixed
constant ($t_{Q}=\frac{\ln10}{\eta}t_{S}$ for $L^{\ast}_{Q}=\eta\Lp$),
and $\alpha\sim0.5$ is weakly dependent on peak luminosity and is
determined by the nearly scale-invariant ``blowout'' of gas as
exponentially growing feedback from black hole growth heats the gas
rapidly and it can no longer cool efficiently in a dynamical time
\citep{H05g}. The lifetime in Equation~(\ref{eqn:qso.lifetime}) is
entirely determined by our simulations, and when quantified as a
function of $\Lp$ in this manner, the quasar lifetime shows no
systematic dependence on host galaxy properties, merger parameters,
initial black hole masses, ISM and gas equations of state and star
formation models, or other varied parameters (see \citet{H05e} for a
detailed discussion and calculation).

Having quantified the luminosity dependent quasar or merger
``lifetime'' in a manner independent of the detailed initial
conditions above, we can map between merger and quasar populations
without requiring any cosmological priors on their distributions. 
Thus while we do not calculate the
distribution of host galaxy properties a priori but rather model
individual mergers in order to achieve the resolution necessary to
model processes related to star formation and black hole growth, 
by quantifying our scalings in terms
of e.g.\ $M_{\rm BH}$ and $\mtot$, our methodology is independent of
the vast majority of host galaxy properties. These differences
manifest in e.g.\ different final stellar or black hole masses, but do
not change any scalings expressed in terms of stellar or black hole
mass. It is straightforward to reverse this procedure, for example starting
with the QLF and quasar lifetime to determine $\nLp$, converting to
$\dot{n}(M_{\ast})$, and using this to predict the MGLF. We can also
compare with other quantities, e.g.\ the SFRF, by convolving
$\dot{n}(M_{\ast})$ with the SFR-dependent lifetime
$t(\dot{M}_{\ast}\,|\,M_{\ast})$.

\subsection{Systematic Uncertainties: Calibrating 
Merger Timescales/Selection Effects and Gas Fractions}
\label{sec:comparisons.systematics}

Before explicitly comparing merger, quasar, and SFR distributions as a
function of redshift, we consider systematic uncertainties introduced
in the various observed samples by e.g.\ selection effects and cosmic
variance.  For quasar populations, these should not be significant
sources of uncertainty, as we generally employ hard X-ray samples
which are complete to luminosities well below those of interest
in our analysis, especially at the moderate redshifts which we
consider in detail \citep[see e.g.][]{Ueda03}, and have been
corrected for obscuring columns with $N_{H}\lesssim10^{25}\,{\rm
cm^{-2}}$.  We have also designed our formulation for comparing e.g.\
quasar and merger populations to be robust with respect to the wide
set of parameters varied in our simulations, as discussed in
\S~\ref{sec:guts}.

However, for a certain ``true'' merger rate as a function of mass,
selection effects will influence the observed merger mass and
luminosity functions.  Of course, there will be some mass/luminosity
completeness limit, but it is straightforward to account for this. 
More problematic is the fact that different definitions of mergers
(i.e.\ choice of maximum separation in pair-selected samples or
strength of tidal features in morphologically identified samples)
will, for the same distribution of merger rates, select different sets
of ``ongoing'' mergers.  Fortunately, we find in \S~\ref{sec:guts}
that merger light curves are quite flat as a function of time, with no
strong dependence of observed total luminosity (in the near-UV through
near-IR bands we consider) on merger stage. This means that the
observed {\em distribution} of merger luminosities and masses will not
be biased by the various sample definitions, only the total number of
objects identified as mergers (i.e.\ vertical normalization of merger
luminosity functions).  This uncertainty is entirely contained within
the parameter $\tmerge$ defined above; i.e.\ the total time over which
a system would be identified as a merger in an observed sample.  In
principle, our simulations can be used to determine this quantity from
first principles for a given observational sample, convolving
predicted galaxy images with the appropriate observational response
(including e.g.\ surface brightness dimming, sky and detector noise)
and selection criteria, to calculate the probability along each
sightline that the object will be identified observationally.

Such an investigation is beyond the scope of the current paper, but is
clearly important for detailed estimates of merger rates as a function
of mass and redshift for a uniformly selected merging galaxy sample at
different redshifts.  Some preliminary results can be obtained by
tracking the path of our simulated mergers in Gini-M20 space, a
non-parametric morphological classification scheme in which spirals,
ellipticals, and mergers occupy distinct regions \citep{Lotz04}.  This
quantitative analysis implies that the galaxy interaction may be
classified as a merger somewhere between $\sim0.5-3.0\,$Gyr, depending
on the viewing angle, the orbit, the progenitor disk orientations, and
the dark matter halo concentration.  The median merger time is
$\approx2.0\,$Gyr and it is worth noting that a similar estimate
obtains from dynamical time considerations and observations
\citep{Patton00}. In what follows, we simply adopt this as the
(constant) $\tmerge$ for our comparison with observed merger
luminosity functions. This is no doubt a rough approximation, but we
can test whether variations in $\tmerge$ will have a significant
impact on our conclusions.

Our mapping between e.g.\ MGLF and QLF is also not {\em entirely}
independent of our simulation parameters, in the sense
that a given QLF defines a unique merger mass distribution, but does
not constrain the initial (pre-merger) gas fractions of the progenitor
galaxies.  For a given mass $\mtot$, the characteristic luminosity at
which the merger is observed can be affected by $\fgas$, at least for
optical/UV samples.  Furthermore, we expect that the distribution of
gas fractions will be drawn from some PDF, which may depend on galaxy
mass. We do not model this (instead adopting a ``mean'' $\fgas$ at a
given redshift), primarily because it introduces additional, poorly
constrained parameters and because it is unclear to what extent this
correlation is driven by more massive systems having undergone
previous major mergers (i.e.\ their pre-merger gas fractions may have
been uniform).  However, the effect is less important than might be
expected, for two reasons.

First, the scaling with mass is not steep (i.e.\ a factor
$\sim10^{2}-10^{3}$ in $M_{\ast}$ yields a factor $\sim2-3$ in
$\fgas$, similar to the existing uncertainty).  Second, our
modeling and the observations of \citet{Xu04,Wolf05,Bundy05a} imply
that the merger mass and luminosity functions do not have steep low
mass/faint-end slopes, meaning that it is a good approximation to
adopt the typical mean ${\langle}\fgas{\rangle}$ of galaxies near the
MGLF ``break''. In detail, we have considered our calculations
adopting the $\fgas$ as a function of late-type galaxy mass from
\citet{RobertsHaynes94}, and find it makes little difference to our
predictions (generally modifying them within existing $\sim1\sigma$
uncertainties).  Of course, to the extent that $\fgas$ evolves with
redshift (see below), this consideration is important in
disentangling how much of that evolution owes to a change in the
characteristic masses of merging galaxies or to gas depletion
through quiescent star formation.

Even if the vertical and horizontal normalizations of the MGLF
(equivalently, $\tmerge$ and $\fgas$) were completely undetermined, we
show in \S~\ref{sec:comparisons.qlf.mglf} that the MGLF and QLF {\em
shapes} are not trivially related, and this alone provides a strong
test of models of quasar lightcurves and triggering.  Moreover, these
systematics (contained in $\fgas$ and $\tmerge$) are reasonably
well-constrained as described above, and the systematic dependencies
in our comparisons are not especially strong over the range in which
they are constrained.

We directly infer the sensitivity of our predictions and comparisons
between e.g.\ the QLF and observed MGLF at a given redshift to these
systematics. In \S~\ref{sec:comparisons} we use the formalism
described in \S~\ref{sec:comparisons.method} and our simulations to
map the observed QLF to a predicted MGLF, and find that it agrees well
with observed MGLFs at $z\sim0.2$ in the $K$-band \citep{Xu04} and
$z\sim0.7$ at 280\,nm \citep{Wolf05}, for expected values of
$\tmerge\sim0.5-3\,$Gyr and $\fgas\sim0.1-0.6$. We can quantify this
formally: for any given ``mean'' $\tmerge$ and $\fgas$ at some
redshift, the QLF and our methodology define a predicted MGLF, and we
can determine the probability (in a $\chi^{2}$ sense) that the
observed MGLF measures this predicted MGLF.

\begin{figure*}
    \centering
    \plotone{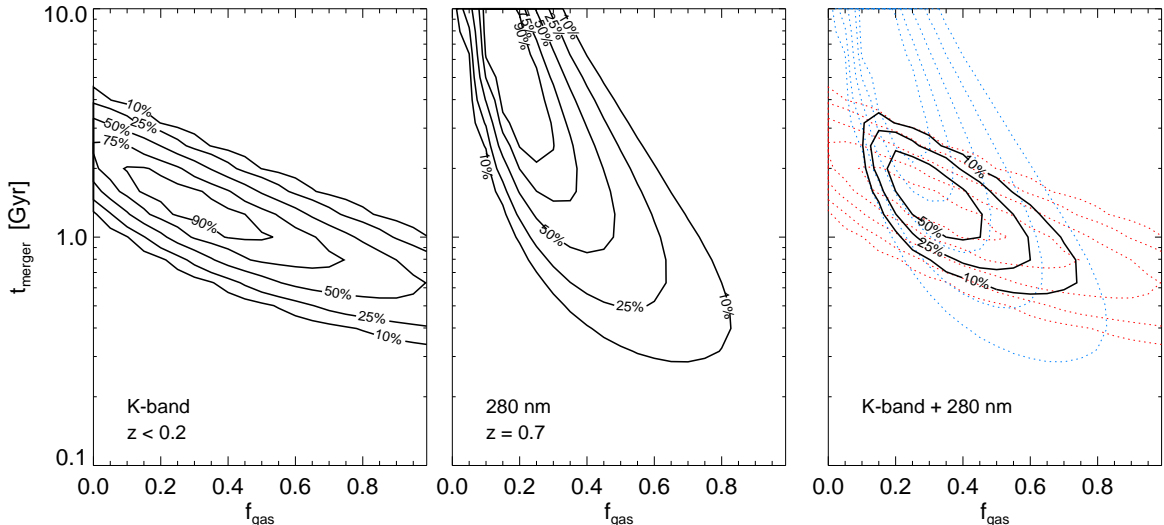}
    \caption{Probability contours in host galaxy gas fraction $\fgas$ and time the 
    merger is visible in the observed sample $\tmerger$ for the MGLF 
    in $K$-band at $z<0.2$ \citep[left;][]{Xu04} and at 280\,nm at $z=0.7$ 
    \citep[middle;][]{Wolf05}. Contours are shown at $10,\ 25,\ 50,\ 75,\ 90\%$ as labeled, 
    and quantify the degree to which systematics will change our results.
    Right panel shows the previous two contour sets (dotted) 
    and the joint (combined) probability contours of both luminosity functions (solid). 
    Allowing these to vary freely yields physically reasonable values, and comparison 
    of the two demonstrates that the dominant systematic uncertainties are different in 
    near-IR and optical/UV MGLFs.     
    Systematics from selection effects and theoretical 
    uncertainties should not change our results over the reasonable ranges of 
    $\tmerger$ and $\fgas$, and will generally be smaller than the effects of 
    cosmic variance.
    \label{fig:P.fgas}}
\end{figure*}

In Figure~\ref{fig:P.fgas}, we plot this for the observed MGLFs at
$z\sim0.2$ and $z\sim0.7$, where probability contours (essentially
measuring the agreement between observed and predicted MGLFs) are
shown in $\tmerge$ and $\fgas$.  In the $K$-band ($z\sim0.2$), this
comparison is essentially independent of $\fgas$ (as expected from the
weak dependence of $K$-band luminosity on $\fgas$ in
Equation~\ref{eqn:K.ML}), and depends only weakly on $\tmerge$ in the
range $\tmerge\sim1-2\,$Gyr.  At $280\,$nm, the dependence on $\fgas$
is stronger, as expected since $L_{280}$ scales with $M_{\ast,\,{\rm
new}}\sim\fgas\,M_{\ast}$, but our results are still essentially
unchanged for reasonable values of typical $z\sim1$ quasar-producing
mergers with gas fractions $\fgas\sim0.2-0.6$ and the broad range of
$\tmerge\sim0.5-3\,$Gyr. Even considering the effects on both MGLFs
simultaneously in the right panel of Figure~\ref{fig:P.fgas} (which
plots the joint probability contours from the $K$-band and $280\,$nm
MGLFs), our comparison is unaffected over the range
$\tmerger\sim1-2\,$Gyr and $\fgas\sim0.2-0.6$ (these numbers should
not be taken too literally, since the samples derive from different
redshifts).

%\clearpage
%\clearpage

In future samples which determine the merging galaxy luminosity
function by uniform, automated selection criteria \citep[see,
e.g.,][]{Lotz06}, our comparisons can be made more precise by
inferring $\tmerge$ a priori from the simulations and directly
measuring $\fgas$ in optical samples.  The key point from the above,
however, is that our theoretical factor $\sim2$ uncertainty in
$\tmerge$ (which stems from not modeling the full sample selection
function and e.g.\ cosmological orbit distributions) is not important
for our conclusions, as any value of $\tmerge$ in this range still
produces a good fit to observed QLFs (i.e.\ gives a self-consistent
mapping between QLF and MGLF).  Likewise, variations in merging galaxy
gas fraction do not change our results.

Moreover, cosmic variance is likely to overwhelm the systematic
normalization issues above for the small volumes probed in observational
measurements of merging galaxy statistics.  We discuss this further in
\S~\ref{sec:evolution.SFR} below, but note that observations with the
deep imaging required to identify merging systems have generally been
limited to small fields.  \citet{Somerville04b} and \citet{Wolf05}
consider this, and estimate that cosmic variance introduces a
systematic factor of $\sim1.5-2$ in the normalization of the merger
luminosity functions.  Therefore, it is not warranted to employ a
more detailed treatment of merger sample selection effects ($\tmerge$) or
$\fgas$ in our modeling, which considers the relation between e.g.\
the MGLF and QLF, if these are measured in different fields and either
(more seriously the MGLF) is significantly affected by cosmic
variance, and in any case the introduced systematic uncertainties are
comparable to current measurement errors in observed MGLFs.

\section{Comparing the Merging Galaxy, Quasar, and 
Star Formation Rate Distributions}
\label{sec:comparisons}

\subsection{The Observed MGLF and QLF}
\label{sec:comparisons.qlf.mglf}

\subsubsection{The $K$-Band Merger Luminosity Function}
\label{sec:comparisons.qlf.mglf.K}

We examine first the observed $K$-band pair luminosity function of
\citet{Xu04}, at low redshift, $z\lesssim0.1$. The pair luminosity
function is determined from the matched 2MASS-2dFGRS 45,289-galaxy
sample of \citet{Cole01}, and agrees well with previous estimates of
the local pair fraction
\citep[e.g.,][]{ZK98,Burkley94,Carlberg94,YE95,Patton97,Patton00} and
the $B$-band luminosity function of paired galaxies
\citep[e.g.,][]{XS91,SR94,KW95,Soares95,Toledo99,Conselice03}, but
accounts for effective pair volume corrections instead of treating
each member singly with its own $V_{\rm max}$ (since both objects must
be identified in a pair to be included in such a sample).  Pairs are
defined to be within a projected separation $\leq 20\,h^{-1}\,{\rm kpc}$,
with velocity difference (where both redshifts are measured)
$<500\,{\rm km\,s^{-1}}$.  Further, the sample is restricted to major
mergers, with $K$-band magnitude differences less than 1\,mag (i.e.\
within a factor $\sim3$ in mass), corresponding well to our modeling
because these mergers are most likely to trigger starburst and quasar
activity and to be visible as peculiar/interacting galaxies for
comparison with morphologically selected MGLFs
\citep{MH94a,Walker96,Conselice03}.  We consider both their binned
data and best-fit Schechter function, with slope
$\alpha=-0.30\pm0.56$, turnover $M_{K,\,\ast}=-23.32\pm0.25$, and
normalization $\log\phi_{\ast}=-3.92\pm0.13$ (converted to our
cosmology).  The distinction between the observed mass function and a
constant merger fraction is only marginally significant; moreover, a
number of studies measure a cumulative merger fraction.  Therefore, we
also consider our predictions for a constant merger fraction as a
function of mass (given the local stellar mass function from
\citet{Bell03}) with the observed low-redshift value $\sim2.5\pm1\%$
\citep{Xu04,Bundy05a,Bundy05b,Bell06}.

%\clearpage
\begin{figure*}
    \centering
    \plotone{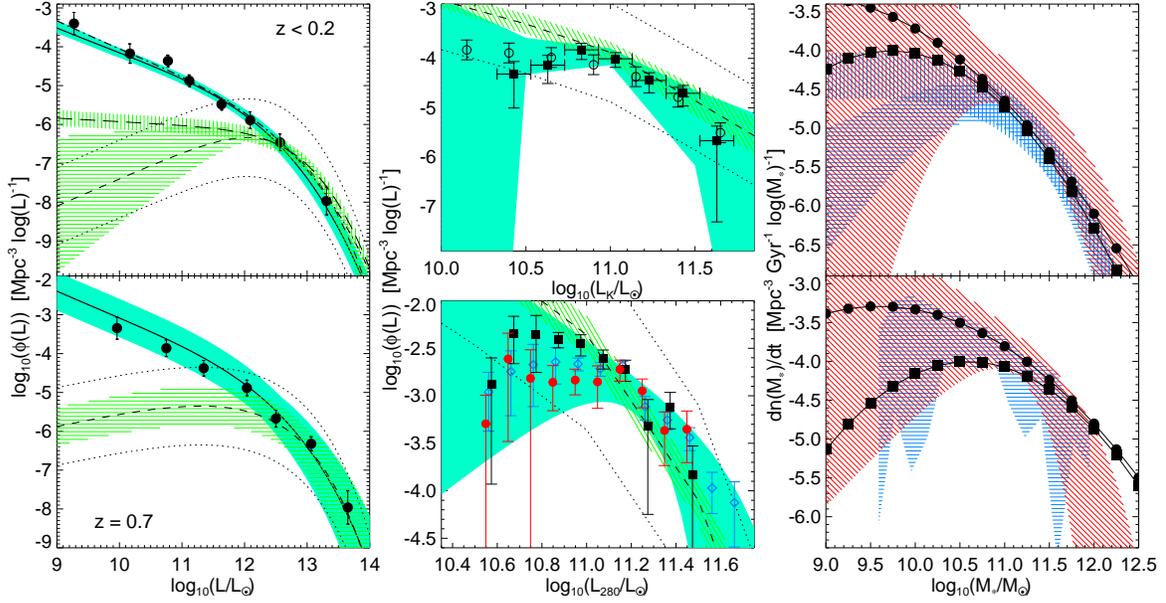}
    \caption{{\it Left}: Predicted QLF from the observed 
    $K$-band $z<0.2$ \citep[upper;][]{Xu04} and 
    280\,nm $z\approx0.7$ \citep[lower;][]{Wolf05} MGLFs, 
    adopting our full model for quasar lifetimes (solid black line; blue shading 
    shows $1\sigma$ range based on errors in the observed MGLF), 
    or a ``light-bulb'' model for the quasar lifetime with $t_{Q}$ 
    varied to produce the best-fit expected QLF (short dashed; green 
    hatched region shows $1\sigma$ range; dotted lines show prediction for $t_{Q}$ a 
    factor of 10 larger or smaller). Circles show the observed hard X-ray quasar luminosity 
    function of \citet{Ueda03}, rescaled to bolometric luminosity with the 
    corrections from \citet{Marconi04}. 
    The predicted QLF from a constant merger fraction $\sim2.5\pm1\%$ at $z<0.2$ is 
    also shown (dot-dashed line uses our full model, long-dashed line a ``light-bulb'' model).
    {\it Middle}: Predicted $K$-band $z<0.2$ and 280\,nm $z\approx0.7$ 
    MGLFs from the QLF of \citet{Ueda03}, 
    lines and colors show the same as the left panels. Observations from \citet{Xu04} from 
    the 2MASS sample 
    are shown as black squares in the upper panel, and a constant merger fraction 
    \citep[$2.5\pm1\%$; given the mass functions of][]{Bell03} as circles. 
    In the lower panel, observations from \citet{Wolf05} are shown 
    as black squares (GOODS), blue diamonds (GEMS), and red circles (GEMS depth on GOODS area).
    {\it Right}: Inferred $\dot{n}(M_{\ast})$, the gas-rich merger rate (spheroid birthrate) with a given stellar mass, 
    from the corresponding ($K$-band, upper; 280\,nm, lower) observed MGLFs 
    ($1\sigma$ range, blue horizontal hatched; instead using a constant merger 
    fraction shown as blue vertical hatched), and the corresponding QLF at 
    $z<0.2$, $z\approx0.7$, respectively ($1\sigma$ range, red shaded). Lines with circles and 
    squares show the fitted $\dot{n}(M_{\ast})$ from the QLF used in \citet{H05g,H05e}, 
    respectively. 
    In all cases, our modeling demonstrates that the observed QLFs and MGLFs are self-consistent 
    and can be used to predict one another, with all mergers producing bright quasars and 
    all bright quasars initially triggered in mergers. Idealized models of the quasar lightcurve 
    give a very misleading relation between MGLF and QLF, predicting a QLF discrepant by 
    $\gtrsim6\,\sigma$.
    \label{fig:LFs}}
\end{figure*}
%\clearpage

Figure~\ref{fig:LFs} shows (upper left panel) the predicted QLF from
the $K$-band low redshift MGLF of \citet{Xu04}. Solid circles indicate
the observed $z<0.2$ hard X-ray QLF of \citet{Ueda03}.
We rescale the hard X-ray QLF to a bolometric QLF for ease
of comparison as in \citet{HRH06}, using a model of the intrinsic quasar continuum SED following
\citet{Marconi04}, based on optical through hard X-ray observations
\citep[e.g.,][]{Elvis94,George98,VB01,Perola02,Telfer02,Ueda03,VBS03}, 
with a hard X-ray reflection component generated by the PEXRAV model \citep{MZ95}. 
This includes the observed dependence of
the optical to X-ray luminosity ratio on luminosity
\citep[e.g.,][]{Wilkes94,Green95,VBS03,RisalitiElvis05,Strateva05}, but for 
our purposes adopting the mean observed quasar SED from 
\citet{Richards06SED} yields a similar result (see
e.g.\ Hopkins et al.\ 2005d, 2006b for a discussion of the impact of different 
bolometric corrections on our predictions). For simplicity, we
consider only the hard X-ray (2-10 keV) QLF, where attenuation 
is usually negligible (and note the \citet{Ueda03} QLF is corrected 
for columns $N_{H}<10^{25}\,{\rm cm^{2}}$).

The black solid line in Figure~\ref{fig:LFs} shows the prediction using 
the quasar lifetimes fitted from our simulations 
(described in \S~\ref{sec:comparisons.method}), 
with the blue shaded region giving the $1\sigma$ range allowed based on the
errors in the observed MGLF (dot-dashed line is the 
result adopting instead a constant merger fraction). We show the results for a characteristic 
$\tmerge=2\,$Gyr, corresponding roughly to the time when
the galaxies are within $\sim50$\,kpc of one another in our
simulations before merging (appropriate for the maximal projected
separation $20\,h^{-1}\approx30\,$kpc of the observed sample), 
and $\fgas=0.15$, corresponding to the typical gas fractions of 
galaxies around the mass of the observed MGLF peak \citep{RobertsHaynes94}. 
However, as shown in 
\S~\ref{sec:comparisons.systematics}, our results are relatively 
insensitive to these choices. Briefly, if we allow $\fgas$ and
$\tmerge$ to vary, a formal best-fit is obtained for
$\fgas=0.25^{+0.21}_{-0.23}$ and $\tmerge=1.7^{+1.2}_{-0.9}$\,Gyr -- in
other words, the quality of the fit is nearly unchanged for the entire
reasonable range of $\fgas$ and $\tmerge$, and these specific values should 
not be taken too literally. The short dashed line shows the prediction
adopting the idealized ``light-bulb'' quasar model described above (\S~\ref{sec:comparisons.method}),
with again the green horizontal cross-hatched range showing the $1\sigma$ range from the
observed MGLF (long-dashed line with vertical hatching shows the 
result for a constant merger fraction). The prediction shown in this case is the best-fit
allowing $\tmerge$, $\fgas$, and the quasar lifetime $t_{Q}$ to vary
freely, giving $t_{Q}\approx6.3\times10^{6}$\,yr for $\tmerge=2$\,Gyr
(only the ratio $t_{Q}/\tmerge$ is constrained) and $\fgas=0.25$. The
dotted lines show the prediction for $t_{Q}$ larger or
smaller by an order of magnitude.

In our model, the faint end of the observed QLF is dominated by
quasars with large peak luminosities, but seen in fainter evolutionary
states during and after the mergers in which they form. Therefore,
the faint end of the predicted QLF is tightly constrained, as it
depends on the MGLF near its turnover, where it is most
well-determined, and there is little difference between adopting the
observed MGLF or a constant merger fraction.  With instead a one-to-one
correspondence between peak and observed quasar luminosity (dashed
lines), the faint-end QLF depends directly on the faint-end MGLF, and
is thus poorly constrained. The agreement between the observed QLF and
our prediction is good, $\reducechi\approx3.41/8=0.43$ and it requires
no fine-tuning of any parameters.  However, even allowing both
horizontal and vertical normalizations to vary freely, the {\em
shapes} alone of the quasar and MGLFs rule out a ``light bulb'' or
exponential lightcurve model at greater than $99.9\%$ confidence, with
the best-fit $\reducechi\approx29.1/7=4.15$ ($\reducechi\approx6.45$
adopting a constant merger fraction). Such a fit furthermore requires
a quasar lifetime $t_{Q}\sim6\times10^{6}\,$yr, at the low extreme of
the range indicated by observations \citep[e.g.,][]{Martini04}.

In any model which reproduces the observed $M_{\rm BH}\propto
M_{\ast}$ relation, $\nLp$ must have approximately the same shape as
the merger rate as a function of $M_{\ast}$, itself given
approximately by the merger mass function.  This shape is
observationally well-defined, with a rapid decline above and
flattening or decrease below the break luminosity. In \citet{H05c}, we
argue that this is the shape of $\nLp$ (equivalently $\dot{n}(\Lp)$)
implied by the combination of our model for quasar lifetimes with the
observed QLF.  However, in idealized models of the quasar lightcurve,
$\nLp$ must have the same shape as the observed QLF, which cannot be
self-consistently resolved with the observed MGLFs. It is true that
the faint-end slope of the MGLF is poorly constrained, with
$\alpha=0.30\pm0.56$; however, to be compatible with the faint end
slope of the QLF in a light-bulb model (i.e.\ to steepen the dashed
line to match the solid line in Figure~\ref{fig:LFs}) requires a
$4\sigma$ change in $\alpha$ ($6\sigma$ relative to a constant merger
fraction), which seems unlikely, given the large quoted error which
includes systematic effects.

Next, we invert this comparison and employ the observed QLF to estimate
the MGLF.  Figure~\ref{fig:LFs} (upper middle panel) shows this, using
the $z<0.2$ hard X-ray \citet{Ueda03} QLF shown in the upper left
panel to determine $\nLp$, and correspondingly $\dot{n}(M_{\ast})$ and
$\phi_{\rm gal}(L_{K})$, the $K$-band $z<0.2$ MGLF.  Again, the 
blue range shows the prediction and $1\sigma$ range (from
errors in the observed QLF) using our full model of quasar lifetimes,
with $\fgas=0.15$, $\tmerge=2\,$Gyr.  Dashed line and green hatched
range shows the same for the ``light-bulb'' quasar lifetime model,
allowing $\fgas$, $\tmerge$, $t_{Q}$ to be fit as in the upper left
panel (producing identical fits), and the dotted lines show the
prediction for an order of magnitude larger or smaller $t_{Q}$. These
can be compared to the black squares, which show the binned MGLF from
\citet{Xu04} (open circles show the constant 
merger fraction estimate). As expected from the upper left panel, the agreement
using our full model of quasar lifetimes is good ($\reducechi=0.35$),
and the agreement with an idealized model of the quasar lifetime is
poor ($\reducechi=3.82$).

The constraints from the observed QLF on the MGLF are
significantly weaker than the constraints on the QLF from the observed
MGLF.  This is not because the QLF is poorly constrained relative to
the MGLF (in fact, the opposite is true). Rather, it is because a given
interval in observed quasar luminosity has significant contributions
from quasars with a wide range of peak luminosities $\Lp\gtrsim L$
(and a correspondingly wide range in host galaxy stellar mass), in
various stages of evolution.  Thus, there are significant degeneracies
in predicting the relative contributions from different 
black hole or host galaxy stellar masses to the
observed QLF based only on its observed shape. This is increasingly
true at fainter merger luminosities, and therefore we can
place only weak constraints on the faint-end $\nLp$ distribution (as
discussed in e.g.\ Lidz et al.\ 2006).  The weak constraints at the
high-$L_{K}$ end, however, owe to larger statistical errors in the QLF.  The wide
range in the allowed faint-$L_{K}$ end of the MGLF 
(demonstrated by the similar predictions given the observed MGLF 
or a constant merger fraction) is reassuring, in
the sense that at low luminosities, selection effects become important
in efforts to identify a sample of merging galaxies, but these will
not significantly change our result. 

As an intermediate stage in these calculations, we have derived the
implied $\dot{n}(M_{\ast})$; i.e.\ the merger rate or birthrate of
spheroids with stellar mass, directly related to the black hole
``triggering'' rate.  We show these $\dot{n}(M_{\ast})$ distributions
in Figure~\ref{fig:LFs} (upper right panel), where the blue horizontal
shaded range shows the $1\sigma$ range implied by the observed MGLF of
\citet{Xu04} (vertically shaded range implied by a constant merger
fraction), and the red range shows the $1\sigma$ range implied by the
observed QLF of \citet{Ueda03} at $z<0.2$. Again, it is clear that the
QLF provides poor constraints on the low-mass behavior of
$\dot{n}(M_{\ast})$. For comparison, we plot the fitted
$\dot{n}(M_{\ast})$ distributions of \citet{H05g,H05e}, which were
fitted to the combination of optical, soft X-ray, and hard X-ray QLFs
\citep{Miyaji01,Ueda03,Croom04,Richards05,HMS05,LaFranca05}.  These
may rise to lower masses before turning over than is implied by the
MGLF, but as we have emphasized above (and shown in e.g.\ Hopkins et
al.\ 2006b) this produces an essentially identical prediction for the
properties of the quasar and remnant red galaxy populations.

\subsubsection{The 280\,nm Merger Luminosity Function}
\label{sec:comparisons.qlf.mglf.280}

We repeat the analysis in \S~\ref{sec:comparisons.qlf.mglf.K}, for the
observed 280\,nm UV MGLF of \citet{Wolf05} from the GEMS and GOODS
surveys at $z\approx0.7$. In detail, the galaxies are selected from
the overlapping area of COMBO-17 and GEMS, with photometric redshifts
$0.65<z<0.75$ and observed $R<24$ (corresponding approximately to a
rest-frame $U$-band).  The sample consists of visually identified
morphologically selected peculiar/interacting galaxies, and as a
consequence there is considerable observational uncertainty in the
faint end of the MGLF, where tidal features are especially difficult
to detect given surface brightness dimming at higher
redshifts. However, as we demonstrate above in the case of the
$K$-band MGLF, our modeling of the QLF is not sensitive to this
faint-end behavior.

Figure~\ref{fig:LFs} shows this comparison in the lower panels, in the
same manner as the upper panels.  The lower left panel shows the
predicted QLF using our full model for the quasar lightcurve (solid
line), and instead for an idealized ``light-bulb'' model for the
quasar lightcurve (dashed line), with uncertainties as in the upper
panels, and dotted lines show the prediction for a fixed quasar
lifetime an order of magnitude larger or smaller.  We compare to the
observed hard X-ray QLF of \citet{Ueda03}, this time at $z=0.7$ (black
circles). Whether we use the GEMS or GOODS data makes little
difference in the prediction of our model, so we show the result using
the GOODS luminosity function, which has a larger number of faint
merging galaxies, maximizing the ability of the ``light bulb'' model
to match the data. Again, it is clear that our model provides a good
fit, while the ``light bulb'' model is ruled out at greater than
$90\%$ confidence ($\reducechi=0.159,\ 2.41$ respectively). The shape
of the QLF alone provides this qualitative constraint on the quasar
lightcurve model, but the vertical and horizontal normalizations also
agree well within the reasonable expected ranges $\tmerge\sim1-3\,$Gyr
and $\fgas\sim0.2-0.6$. As above, $\tmerge$, $\fgas$, and $t_{Q}$ are
allowed to vary freely in the ``light bulb'' model, and although the
fit is poor, the best-fit values are $\fgas=0.3$ and
$t_{Q}=5.8\times10^{6}$\,yr for $\tmerge=2$\,Gyr.  A
$\sim4\sigma$ change in the faint-end slope of the MGLF would be
required for the ``light bulb'' model to agree with the QLF, even for
the maximal estimate of the number of faint mergers in \citet{Wolf05}.

The lower middle panel of Figure~\ref{fig:LFs} shows 
the 280\,nm MGLF inferred from the $z=0.7$ QLF of
\citet{Ueda03}, in the same style as the upper middle
panel. The inferred MGLFs can be compared to those of \citet{Wolf05}
from the GOODS survey, GEMS survey,
and a combination of GEMS depth on the GOODS area.  The
differences in the observed MGLFs at the faint end demonstrate the
considerable observational uncertainty here, but the
QLF is not sensitive to this faint-end behavior. This is not the case
for the idealized ``light bulb'' model, as demonstrated by the narrow
range of the green hatched region even at low luminosities. The constraints are 
weaker, but the agreement between our 
model predictions and the observed MGLF is good ($\reducechi=0.31$,
compared to $\reducechi=2.15$ for the light-bulb model).

Finally, we show the $\dot{n}(M_{\ast})$ distribution implied by the
MGLF of \citet{Wolf05} (lower right panel) as the blue shaded range,
The range implied by the QLF at $z=0.7$ is indicated as the red shaded
range, with the fits to $\dot{n}(M_{\ast})$ used in \citet{H05g,H05e}.
The implied $\dot{n}(M_{\ast})$ distributions at $z=0.7$ and $z<0.2$
from 280\,nm and $K$-band observations, respectively, can be compared
to infer the evolution of merger rates and quasar birthrates as a
function of mass, but we caution against taking this comparison too
far. The samples use various selection criteria, are from different
surveys, are not in the same wavebands, and have different dependence
on gas fraction. Therefore, the potential systematic effects may be
quite different between the two.

\subsubsection{Merger Mass Functions}
\label{sec:comparisons.qlf.mglf.MFs}

We repeat our analysis using the morphologically separated 
mass functions of \citet{Bundy05a,Bundy05b} from GOODS and DEEP2. Stellar 
masses are determined by fitting to optical spectroscopy and near-IR 
photometry, and we adopt the mass functions for objects visually classified as 
mergers \citep[following][]{Brinchmann98} with ACS morphologies. This gives a 
binned merger stellar mass function determined in each of the intervals 
$z=0.2-0.55$, $z=0.55-0.8$, and $z=0.8-1.4$, spanning 
$M_{\ast}\sim10^{9}-10^{12}\,M_{\sun}$
(see Figure~\ref{fig:compare.mf.lf} below for a detailed comparison 
with the $z=0.55-0.8$ mass function). In each case, the observed merger mass function 
implies a QLF in good agreement with the observations of \citet{Ueda03}
($\reducechi\approx 0.85,\ 0.52,\ 0.98$
%6.0/7,\ 3.1/6,\ 4.9/5$ 
for $z=0.2-0.55,\ 0.55-0.8,\ 0.8-1.4$, respectively), and the implied 
mass function from the QLF, albeit less well-constrained, is consistent with the 
observed mass functions. These observations also consistently rule out 
a ``light bulb'' or similarly idealized model for the quasar lightcurve at $\gtrsim10\sigma$.

\subsection{Comparison of Merger Mass and Luminosity Functions}
\label{sec:comparisons.mf.lf}

While the agreement between quasar and merger statistics is encouraging, 
we now test our modeling of the MGLF directly by 
comparing the observed 280\,nm MGLF of \citet{Wolf05} to the 
measured merger mass functions of \citet{Bundy05a,Bundy05b}. 
We consider the \citet{Bundy05a} mass functions determined from the GOODS field, 
with redshifts $0.55<z<0.8$, similar to the $0.65<z<0.75$ range considered 
in \citet{Wolf05}. In both cases, optical or near-IR morphologies are used to visually select merging 
systems, so the ambiguity in selection effects is minimized. 

\begin{figure}
    \scalesinglefig
    \centering
    \plotone{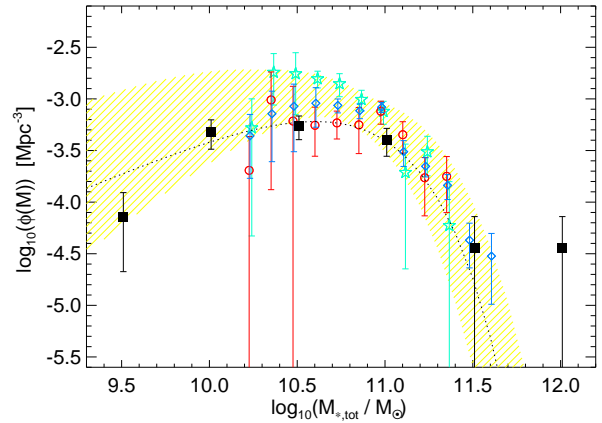}
    \caption{Merger mass function at $z\approx0.7$, 
    predicted from our modeling with the observed 280\,nm MGLF of 
    \citet{Wolf05} (colored), and directly calculated from 
    near-IR and optical observations in 
    \citet{Bundy05a} (black). 
    Points converted from the 280\,nm observations in Figure~\ref{fig:LFs}
    are shown as cyan stars (GOODS), blue diamonds (GEMS), and
    red circles (GEMS depth on GOODS area), and the $\sim1\sigma$ 
    allowed mass function from our modeling as the yellow shaded region. 
    Black squares show the binned mass function from 
    \citet{Bundy05a}, black dotted line shows the best-fit Schechter function. 
    The agreement implies that our simulations 
    reliably map between MGLFs and merger stellar or 
    black hole mass functions. 
    \label{fig:compare.mf.lf}}
\end{figure}

Figure~\ref{fig:compare.mf.lf} shows the 
inferred merger mass function using our simulated $P(L_{280}\,|\,M_{\ast})$ 
to convert $\phi(L_{280})$ directly to $\phi(M_{\ast})$. 
The observations of 
\citet{Wolf05}, converted to a mass function, are shown by the symbols
indicated in the caption.  Alternatively, adopting the $\sim1\sigma$ range in allowed 
fitted MGLFs from \citet{Wolf05} and converting this to a merger mass 
function yields the shaded yellow range. The range shown can also be 
thought of as a range in gas fraction, with $f_{\rm gas}\approx 0.15 - 0.30$.
The observed merger mass function of \citet{Bundy05a} is shown as the 
black squares, with the black dotted line showing the best-fit Schechter function. 
%\clearpage
%\clearpage

The merger mass function implied by the MGLF and our modeling 
agrees well with that observed, suggesting that we are capturing the 
critical dependence of observed merger luminosities on host properties. 
Coupled with observations of a similar relation between black hole and spheroid mass 
at these redshifts \citep[e.g.,][]{Peng06}, this observationally supports
multiple implicit levels of our modeling. 

\subsection{The Distribution of Star Formation Rates}
\label{sec:comparisons.sfr}
%\clearpage
\begin{figure}
    \scalesinglefig
    \centering
    \plotone{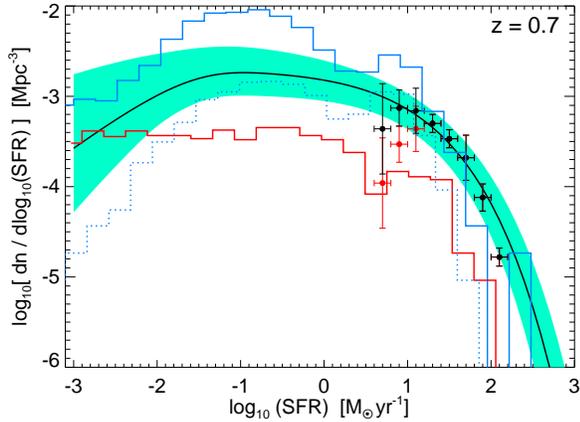}
    \caption{Distribution of star formation rates in mergers at $z=0.7$, 
    inferred from our modeling and the $z=0.7$, 280\,nm MGLF of \citet{Wolf05}
    (solid line). Shaded range shows $1\sigma$ range based on observational errors 
    in the MGLF. Circles show the observed SFRF of \citet{Bell05} at the 
    same redshift (black, corrected for completeness; red, uncorrected), normalized 
    with the observed IR luminosity functions of \citet{LeFloch05}. 
    Red line shows the predicted SFRF owing to major mergers 
    from the semi-analytic models of \citet{Somerville01}, blue lines 
    include minor mergers. Solid lines adopt $\tmerge\sim10\,t_{\rm dyn}\sim\,$Gyr, 
    dotted lines $t_{\rm dyn}\sim\,10^{8}\,$yr as the maximum time since the 
    last merger. The 
    agreement implies that our modeling reliably predicts the 
    merger-driven SFR distribution from a given MGLF or QLF. Note the 
    increasing importance of minor mergers at low SFR. 
    \label{fig:sfrf.z07}}
\end{figure}
%\clearpage

Having determined $\dot{n}(M_{\ast})$ from the QLF or MGLF, we can
convolve instead with the time each merger spends in some interval in
SFR, $\sfr$ (Equation~[\ref{eqn:t.sfr}]), to determine the observed
SFRF in mergers. Figure~\ref{fig:sfrf.z07} shows the results given
$\dot{n}(M_{\ast})$ (specifically $\dot{n}(\mnew$)) determined from the
280\,nm observed peculiar/interacting luminosity function of
\citet{Wolf05} at $z=0.7$ from GEMS and GOODS (solid line with
$1\sigma$ blue shaded range). We compare this to the observed SFRF of
the same merging galaxy sample, determined in \citet{Bell05}.
Note that while \citet{Bell05} determine the SFRF and
relative number density as a function of morphology, they do not
determine the absolute SFRF normalization (i.e.\ proper effective
volume correction), and so we adopt that inferred by \citet{LeFloch05}
at $z=0.6-0.8$ from Spitzer in the Chandra Deep Field South. Red
points are observations with no correction for
incompleteness below $\sfr\sim10\,M_{\sun}{\rm yr^{-1}}$, while
black
points have been corrected for incompleteness at low SFR based on the
comparison of GEMS and GOODS luminosity functions from \citet{Wolf05}.

The agreement is good, implying that we are
properly modeling the relation between the optical/UV luminosity of
mergers and at least the statistics of the induced star formation as a
function of time. This is not trivial, as 
there is no one-to-one correspondence between the 280\,nm
luminosity and SFR (as optical/UV lightcurves are flat while 
SFRs change by orders of magnitude). 
Also note that, because both the 280\,nm
magnitude and SFR distributions are functions of $\mnew$ (as opposed
to $\mtot$), the relation between the two is independent of $\mtot$
and $\fgas$.

It is also of interest to compare our prediction for the
distribution of mergers and SFRFs to that
from semi-analytic models.  In general, this is difficult, as the
events of interest are not mergers of dark matter halos, but involve
the luminous galaxies themselves, in which the galaxies are of
comparable mass and have a large supply of cold (rotationally
supported) gas. Furthermore, the exact requirements for triggering
starbursts and AGN depend on, for example, the pressurization of ISM
gas, star formation recipes, and the distribution of orbital
parameters in the mergers (e.g., Hernquist \& Mihos 1995).  We have
specifically chosen to quantify the properties of mergers in terms of
e.g.\ $\Lp$, $\mtot$, and $\mnew$, in a manner which suppresses these
dependencies, in order to relate merger, starburst, and quasar
populations independent of these detailed (and difficult to model)
cosmological distributions. It is therefore outside the scope of this
paper to consider a detailed comparison of merger rates as a function
of these conditions as implied by e.g.\ our semi-empirical modeling,
semi-analytic models, and cosmological simulations. However, we
briefly consider a comparison between our predicted merger-driven SFRF
and that predicted by semi-analytic models in
Figure~\ref{fig:sfrf.z07}.

We show as the red solid line the SFRF predicted owing to major
mergers (mass ratios less than $\sim$3:1), and as the blue solid line 
the SFRF owing to ``minor'' mergers (mass ratios less than
$\sim$10:1), calculated at $z=0.7$ from the semi-analytic models
of \citet{Somerville01,Somerville04a}. We consider a galaxy to 
be ``merging'' if it has undergone a merger within a 
time $\sim10\,t_{\rm dyn}\sim$Gyr (blue dotted line shows 
the minor-merger prediction for a merger time $\sim t_{\rm dyn}\sim 10^{8}\,$yr). 
There is broad agreement, suggesting that the rate of gas-rich mergers in the
standard CDM cosmology is appropriate to that implied by our modeling
and the observations. However, it is only with the
inclusion of some ``minor'' mergers that the semi-analytic model
predictions match the observations and our predictions from the QLF
and MGLF, suggesting that our implied merger rates as a function of
final, total stellar mass may have contributions from a wide range of mass
ratios $\sim$10:1 to 1:1 (which we implicitly account for by
expressing the merger rate in terms of the final mass and peak quasar
luminosity), increasingly dominated by major mergers at the high-mass
end. This comparison, though, is sensitive to the 
estimate of the observable timescales in the semi-analytic model 
\citep[but see e.g.,][for comparison with these merger rate estimates]{Bell06}.

\section{The Evolution of Merger Luminosity Functions and 
Star Formation Rate Distributions}
\label{sec:evolution}

\subsection{Merger Luminosity Functions}
\label{sec:evolution.MGLF}

Now, we use the observed QLF to predict the MGLF at 
redshifts where measurements do not yet exist. We
have seen that the observed QLF is insensitive to the number of 
low-mass mergers. If we assume that the MGLF is well-fitted 
by a Schechter function,
\begin{equation}
\phi(M)=\phi_{\ast}\,10^{0.4\,(M-M_{\ast})\,(\alpha+1)}\exp{\bigl\{}-10^{0.4\,(M-M_{\ast})}{\bigr\}}, 
\end{equation}
this means that the QLF has almost no power to constrain
$\alpha$. However, if we assume a specific $\alpha$, we can
then fit the observed QLF to determine the allowed range of
$\phi_{\ast}$ and $M_{\ast}$.

%\clearpage
\begin{figure}
    \scalesinglefig
    \centering
    \plotone{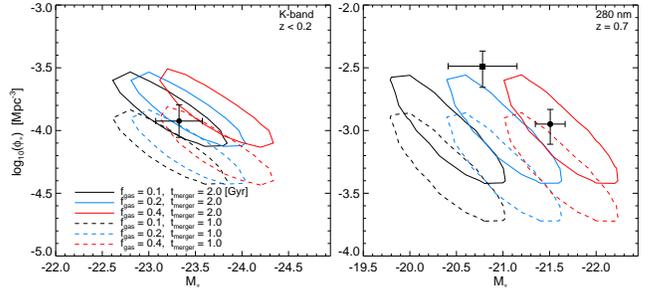}
    \caption{Contours of the $1\sigma$ allowed Schechter function parameters 
    $M_{\ast}$, $\phi_{\ast}$ (assuming $\alpha=-0.5$) for the MGLF 
    in $K$-band at $z<0.2$ (left) and at 280\,nm at $z=0.7$ (right), 
    assuming different values for the gas fraction $\fgas$ and selection efficiency 
    (time during which the merger will be identified as such) $\tmerge$, as labeled.
    Points with error bars show the measurements at the corresponding frequency and redshift 
    of \citet{Xu04} from the 2MASS-2dFGRS survey in $K$-band and 
    \citet{Wolf05} from GOODS (square) and GEMS (circle) at 280\,nm
    (observations rescaled assuming $\alpha=-0.5$). The QLF can significantly 
    constrain the allowed MGLF parameters $\phi_{\ast}$ and $M_{\ast}$, 
    modulo systematic uncertainty in the merger timescale $\tmerge$ (i.e.\ sample 
    selection effects) and typical merging galaxy gas fraction $\fgas$.
    \label{fig:P.mstar}}
\end{figure}
%\clearpage

Figure~\ref{fig:P.mstar} shows this determination of $\phi_{\ast}$ and
$M_{\ast}$, assuming a constant $\alpha=-0.5$, for both the $K$-band
$z<0.2$ and 280\,nm $z=0.7$ MGLFs, compared to the observations of
\citet{Xu04} and \citet{Wolf05}. The observed MGLF
$\phi_{\ast}$ and $M_{\ast}$ have been re-fit assuming the same
$\alpha=-0.5$. For each
wavelength, we show the contours enclosing the $1\sigma$ range of
$\phi_{\ast}$, $M_{\ast}$ for a given gas fraction and
$\tmerge$. As expected from \S~\ref{sec:comparisons.systematics}, the 
theoretical uncertainty in $\tmerge$, $\fgas$ is generally unimportant 
relative to the observational factor $\sim2$ uncertainties. 
In principle, the number of observed mergers scales 
with the selection efficiency $\tmerge$ and their luminosities with $\fgas$, 
such that they independently control the
prediction of $\phi_{\ast}$ and $M_{\ast}$, respectively. If there
were no errors in observations of the relevant luminosity functions,
comparison of $M_{\ast}$ alone could fix the distribution of $\fgas$. However, even
small errors in the observed luminosity functions make $\phi_{\ast}$
and $M_{\ast}$ strongly degenerate parameters, and the choice 
of $\alpha$ introduces another, albeit well-known, degeneracy. 

%\clearpage
\begin{figure*}
    \centering
    \plotone{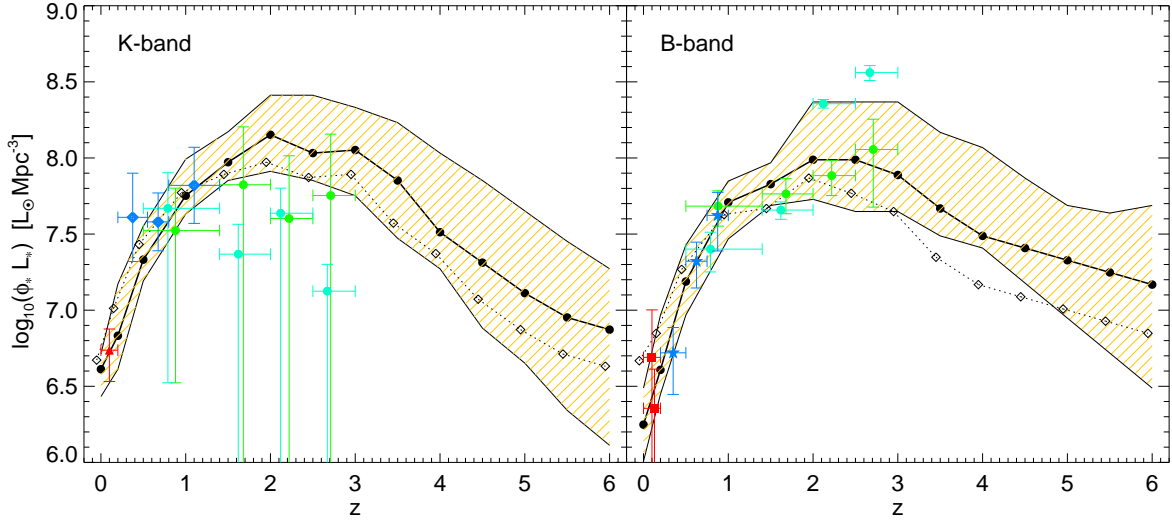}
    \caption{Predicted luminosity density contributed by merging galaxies 
    in $K$-band (left) and $B$-band (right), 
    as a function of redshift. Solid line assumes a gas fraction $\fgas=10^{-t/t_{H}}$, 
    dotted line assumes a constant $\fgas=0.3$ at all $z$. Yellow shaded range 
    shows the $1\sigma$ allowed range inferred from the QLF, for the 
    evolving $\fgas$ case (errors are similar for constant $\fgas$).
    Observations are shown from 
    \citet{Xu04} (red triangle), \citet{Bundy05a,Bundy05b} (blue diamonds), 
    \citet{Conselice03,Conselice05} (circles; HDF-N in cyan, HDF-S in green), 
    \citet{Toledo99} (red squares), and \citet{Brinchmann98} (blue stars). 
    The observed mass/luminosity density in mergers and its evolution is consistent with that required 
    if all bright quasars are triggered in mergers {\em and} the converse, that 
    all mergers trigger bright quasars. The quasar luminosity function and our modeling 
    can be used to predict the merger mass and luminosity density at 
    all redshifts where the QLF break is reasonably constrained. 
    \label{fig:lum.density}}
\end{figure*}
%\clearpage

Given these degeneracies, we focus on the more robust prediction of the 
luminosity density. The full luminosity density,
$\rho_{L}=\phi_{\ast}\,L_{\ast}\,\Gamma(\alpha+2)$, technically
includes the order unity correction from the integrated faint-end
contribution, $\Gamma(\alpha+2)$, which our modeling has little power
to constrain from the observed QLF.  However, the quantity
$\phi_{\ast}\,L_{\ast}$ is well-determined, and in
Figure~\ref{fig:lum.density} we show our predictions for this
combination (i.e.\ neglecting the $\Gamma(\alpha+2)$ correction, or
equivalently assuming $\alpha=-1$ or $\alpha=0$, although the
corrections from $\alpha$ are small for the reasonable range
$-1.5\leq\alpha\leq0.5$). Most observational estimates of 
$\rho_{L}$ have been in the $B$-band, so we show the predicted luminosity density in
$K$-band (left) and $B$-band (right), as a function of redshift from
$z=0-6$, based on the combined 
\citet{Miyaji01,Ueda03,Croom04,Richards05,HMS05,LaFranca05} QLFs. 
The $B$-band (440\,nm) 
calculation of host galaxy luminosities is identical to our 
calculation at 280\,nm \citep[see also][for a full calculation]{H05e}; 
the luminosities and scaling with $\mnew$ are nearly identical 
(see Table~\ref{tbl:magslookup}) 
modulo a nearly mass-independent $M_{B}-M_{280}\sim0.9$ (reflecting the 
mean young stellar population age $\sim0.5\,$Gyr). 
We compare with the $z<0.2$ MGLF measurement of
\citet{Xu04}, the merger IR and $B$-band 
luminosity densities of \citet{Conselice03,Conselice05} at mean redshifts
$z=0.88,\ 1.68,\ 2.22,\ 2.71$,
the peculiar/interacting galaxy mass functions of 
\citet{Bundy05a,Bundy05b} at $z=0.2-0.55$, 
$z=0.55-0.8$, and $z=0.8-1.4$, the 
$z<0.1$ optical pair luminosity function of
\citet{Toledo99}, and $B$-band peculiar
luminosity densities from \citet{Brinchmann98}
at $z=0.2-0.5$, $z=0.5-0.75$, and $z=0.75-1.0$.

In order to predict $\rho_{L}$ at each redshift from the QLF, we must
adopt a typical gas fraction $\fgas(z)$. In Figure~\ref{fig:lum.density} the 
dotted line is for a constant $\fgas=0.3$, where this value is chosen because it provides
a best fit to the cumulative observations plotted (considering the fit to the 
full MGLF where possible). The solid line 
assumes an exponentially declining gas fraction with cosmic time,
$\fgas=10^{-t/t_{\rm H}}$ where $t_{\rm H}$ is the Hubble time, such
that $\fgas=1$ at early times and $\fgas=0.1$, similar to the Milky
Way, at present. This gives an $e$-folding time for $\fgas$ of $t_{\rm
H}/\ln(10)\approx6\,$Gyr, similar to that expected for quiescent
spirals following a Schmidt-type star formation law
\citep{Kennicutt98,RowndYoung99,MartinKennicutt01,SH03,Li05a,Li05b}. 
Comparing with the evolution of the distribution of cold
(rotationally supported) gas mass fractions in galaxy mergers (with
total baryonic mass above $10^{10}\,M_{\sun}$) in the
semi-analytic models of \citet{Somerville04a}, we find that these
cosmological models predict a similar evolution in the mean gas
fraction. The yellow shaded range shows the $1\sigma$
allowed range inferred from the QLF, for the evolving $\fgas$
case (uncertainties are similar for constant $\fgas$). 

Although it is clear that the observations do not strongly distinguish
between the two cases, and we caution that we have assumed a constant
selection efficiency $\tmerge$ despite different observed sample
selection criteria and different redshifts, there is a significantly
better fit to the combination of luminosity density and MGLF observations 
in the case of the exponentially declining gas fraction
($\reducechi=0.4$ compared to $\reducechi=2.1$). This distinction is
derived from the $B$-band and 280\,nm data; the $K$-band is
relatively insensitive to $\fgas$ and does not significantly
constrain this evolution. That the exponential decline in gas fraction
is preferred by the data is expected, and the observations are not
accurate enough to significantly constrain the timescale or detailed
functional form of the decrease in gas fraction with time, but this
demonstrates the key qualitative behavior, that merging galaxies were
more gas rich in the past and that these measurements can constrain
that evolution through the evolution in $M_{\ast}(z)$.

It is clear from the figure that
our predictions for the luminosity density are narrowly constrained,
generally within a factor $\sim2$ at all redshifts, with the errors 
dominated by uncertainties in the observed QLF (and thus potentially improved
by future, high redshift complete samples which can constrain the QLF
break luminosity at a range of redshifts) as opposed to fitting
degeneracies. The variation owing to different gas fractions is small
in the $K$-band, as expected, but enters approximately linearly into the
$B$-band luminosity densities. Improved constraints on the $K$-band
MGLF measurements and extension of $B$-band MGLF measurements to
higher redshifts can provide a strong test of whether the consistent
relation between merger and quasar populations this modeling
demonstrates at $z\lesssim3$ remains true at high redshifts.

\subsection{The Merger-Driven Star Formation Rate Distribution}
\label{sec:evolution.SFR}

%\clearpage
\begin{figure*}
    \centering
    \plotone{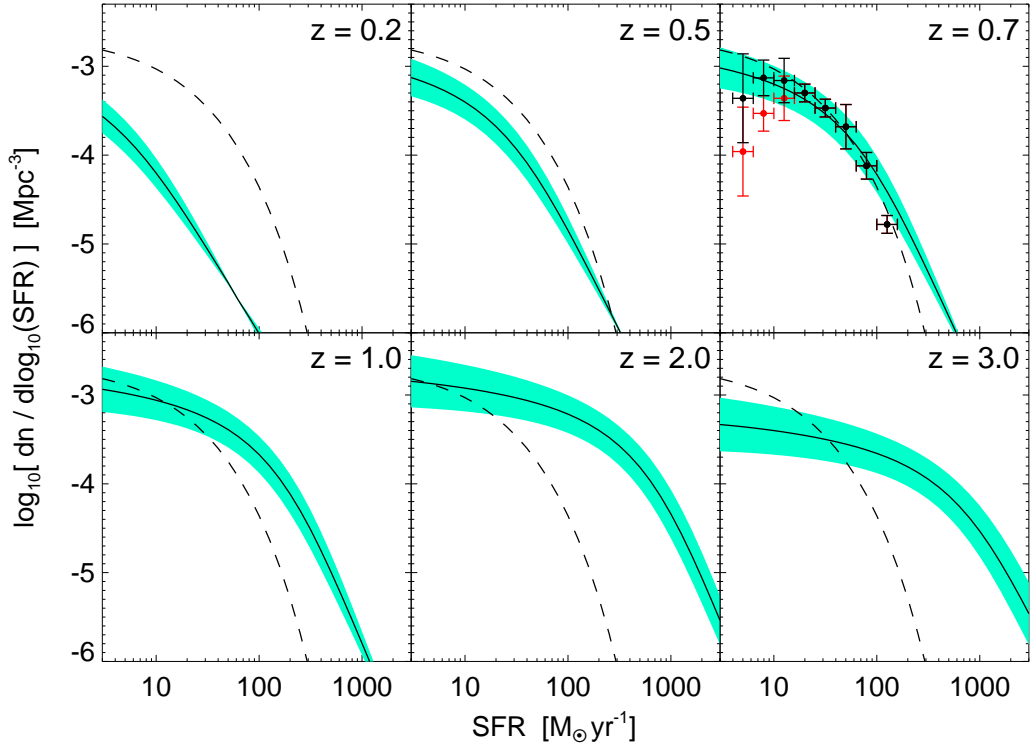}
    \caption{Inferred SFRF in mergers from the QLF as a function of redshift (black lines) at 
    the redshifts shown. The shaded area shows the $1\sigma$ range from errors in the observed 
    QLF and uncertainties in fitting to it. Observations of \citet{Bell05,LeFloch05} 
    are shown at $z=0.7$ in the manner of Figure~\ref{fig:sfrf.z07}, and the dashed 
    line in each panel shows the prediction from the 280\,nm MGLF of \citet{Wolf05} 
    from Figure~\ref{fig:sfrf.z07} at $z=0.7$ for comparison. The dashed line and 
    solid line at $z=0.7$ are slightly different because the former is predicted from the 
    observed merger luminosity function at that redshift, the latter from the observed 
    QLF. The QLF alone allows us to reasonably constrain the distribution of star 
    formation rates in mergers, over a wide range of redshifts where observations 
    are not available. 
    \label{fig:sfrf}}
\end{figure*}
%\clearpage

Following \S~\ref{sec:comparisons.sfr} but beginning from the observed QLF instead 
of the MGLF, we can predict the SFRF as a function of redshift. 
Figure~\ref{fig:sfrf} shows the results of this
calculation at several redshifts $z=0-3$ (as labeled), 
from the fit to the 
combined \citet{Miyaji01,Ueda03,Croom04,Richards05,HMS05,LaFranca05}
QLFs. The dashed line shows the prediction from the MGLF (as opposed to the QLF)
at $z=0.7$ from Figure~\ref{fig:sfrf.z07}. 
Points at $z=0.7$ show the observations
\citep{Bell05,LeFloch05} as in Figure~\ref{fig:sfrf.z07}. 
We adopt our rough estimate 
$\fgas=10^{-t/t_{H}}\approx\exp(-t/6\,{\rm Gyr})$. 
The evolution in the SFRF is predominantly a reflection of the luminosity
evolution of the QLF, as the break in the QLF evolves towards 
larger luminosities with redshift, implying that typical corresponding black hole 
and host galaxy masses must evolve similarly. Our assumption that 
the typical gas fractions rise with redshift also 
increases the SFR for a given total (final) mass, resulting in a
further factor $\sim3$ increase in $\mnew$ from $z=0.2-2$. 
Comparing e.g.\ the specific star formation rate distributions as a function 
of stellar mass, age, and redshift from our simulations and observed in 
\citet{Bauer05,Feulner05,PerezGonzalez05,Papovich06} 
yields a similar self-consistent ``downsizing.''

Although this calculation suggests a form for e.g.\ the ULIRG
and LIRG IR luminosity functions, we caution that we do not model 
re-radiation of absorbed light by dust and are not yet in a position to predict 
these properties. However, a rough calculation shows that we expect
ULIRGs counts to be consistent with our modeling \citep{H05e}. We
defer a more thorough calculation of the IR spectrum, including
dust heating, re-radiation and line emission, to future
work.

%\clearpage
\begin{figure*}
    \centering
    \plotone{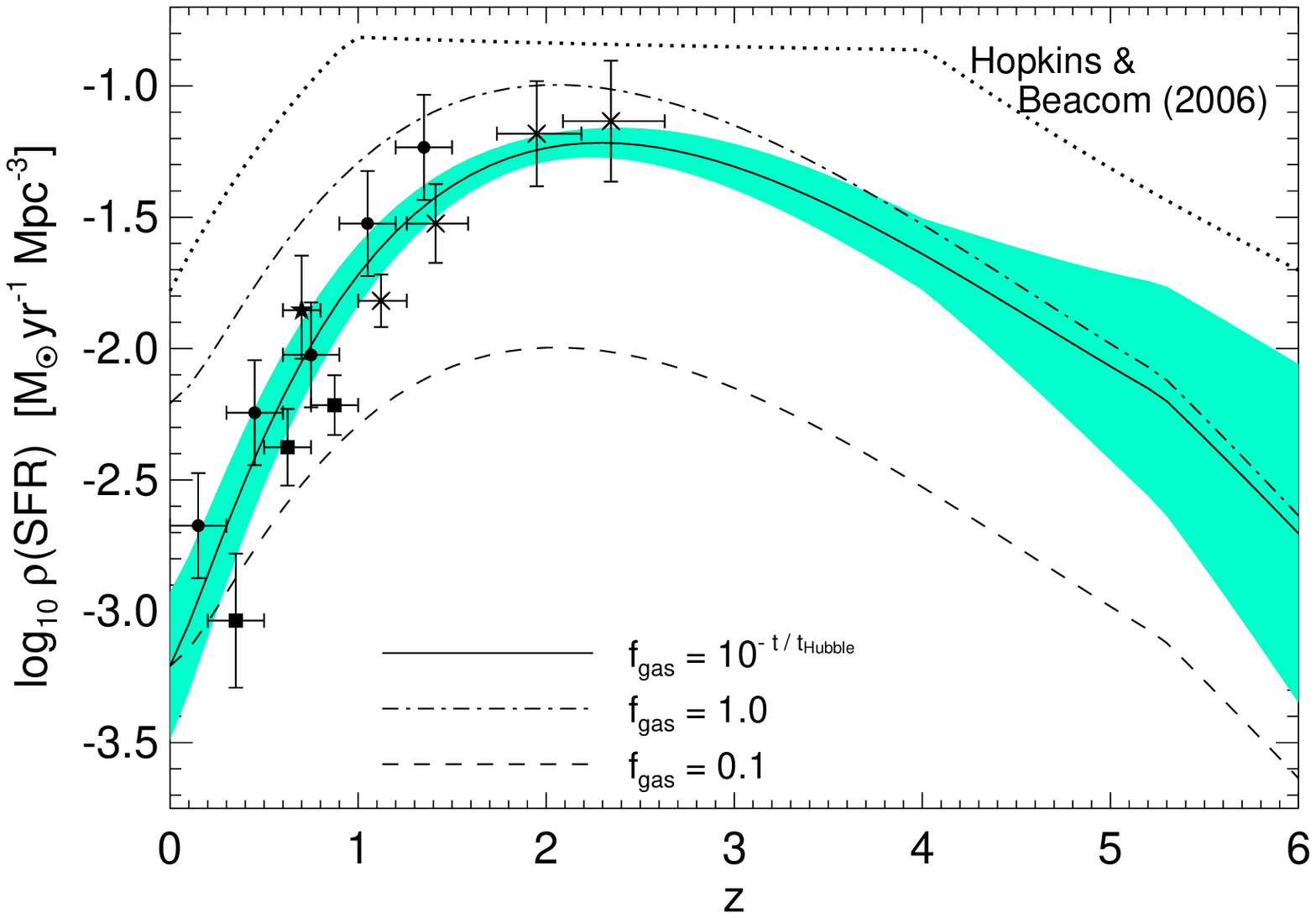}
    \caption{Star formation rate density in mergers, inferred from the quasar luminosity 
    function assuming an average gas fraction $\fgas=10^{-t/t_{H}}=\exp(-t/6\,{\rm Gyr})$
    (solid line). The shaded area shows the $1\sigma$ range allowed 
    (owing to degeneracies in fitting
    to the observed QLF). Note that adopting the steeper faint-end slope estimates from the 
    deeper GOODS subsample of \citet{Wolf05} systematically 
    increases our prediction by a 
    factor $\sim50\%$ (at the upper range of the $1\sigma$ uncertainty).
    The dot-dashed line and dashed lines show the result assuming a 
    constant gas fraction $\fgas=1.0$ and $\fgas=0.1$, respectively (for clarity, 
    errors are not shown, but they have identical form). Dotted lines show the 
    estimated total (extinction-corrected) star formation rate density from 
    \citet{HopkinsBeacom06}. 
    Observations of the SFR density in mergers at various redshifts 
    are shown from \citet{Bell05} (star), \citet{Brinchmann98} (squares), 
    \citet{PerezGonzalez05} ($\times$'s; from the estimate of the 
    ULIRG contribution to the SFR density), 
    and \citet{Menanteau01,Menanteau05} (filled circles). The $\fgas=1$ line 
    can also be thought of as the total rate at which stellar mass is ``moved'' or ``generated''  
    onto the red sequence by quasar-producing gas-rich mergers. 
    The QLF alone can be used as an independent constraint on the merger-driven SFR 
    density of the Universe; the SFR density in mergers must rise from $\sim1\%$ 
    of the total SFR density at $z=0$ to $\sim15-25\%$ at $z=1$ and 
    $\sim30-50\%$ at $z=2$. 
    \label{fig:sfr.int}}
\end{figure*}
%\clearpage

We can integrate the SFRF calculated at each redshift to determined the SFR
density in mergers as a function of redshift, as shown in Figure~\ref{fig:sfr.int}.
The solid black line is our prediction 
assuming $\fgas=10^{-t/t_{H}}$
as estimated in \S~\ref{sec:evolution.MGLF}, with shaded $1\sigma$
range based on the errors in the observed QLF and 
degeneracies in fitting. We also show
the predictions for a constant $\fgas=1.0$ (dot-dashed) or
$\fgas=0.1$ (dashed) (errors similar to the case shown). 
The solid line
effectively interpolates between a high-gas fraction era at
$z\gtrsim2$ to present characteristic low gas fractions $\fgas\sim0.1$ at $z=0$. 
The predicted evolution in $\rho({\rm SFR})$ is understood from
the QLF.
The time-averaged growth rate of black holes is $j_{L}/\epsilon_{r}\,c^{2}$, 
where $j_{L}$ is the quasar luminosity density, so if all bright quasars 
(i.e.\ most of $j_{L}$ at moderate/high redshift) come from spheroid-forming 
mergers with $\mnew\approx\fgas\,M_{\ast}\approx \fgas M_{\rm BH}/\mu$ 
($\mu\approx0.001$) we expect 
$\rho({\rm SFR})\sim\fgas\,j_{L}/\mu\,\epsilon_{r}\,c^{2}$. 

For comparison, we show the total (integrated over all
morphological types), extinction-corrected SFR density estimated in
\citet{HopkinsBeacom06} (similar to e.g. Cole et al. 2001, Hopkins 2004
or predictions of cosmological simulations, e.g. Springel \&
Hernquist 2003b, Hernquist \& Springel 2003) as the dotted line. 
Observations of the SFR density in merging/peculiar systems are
plotted, from \citet{Bell05}, \citet{Brinchmann98},
and
\citet{Menanteau01,Menanteau05}. The 
estimate of the SFR density in ULIRGs from \citet{PerezGonzalez05} 
is also shown, but we caution that while these 
objects are associated with mergers at low redshift, 
dusty, more concentrated disks at high redshift may contribute
to this population. 

The agreement between the observations of the SFR density in merging galaxies as a
function of redshift and our modeling is good,
implying that the QLF can be used as an independent, albeit
indirect constraint on the star formation history in mergers and the
contribution of merger-driven starbursts to present stellar
populations. Systematic uncertainties in the observations are still a 
concern; e.g.\ the 
observations of \citet{Brinchmann98} may be incomplete 
owing to dust extinction; see \citet{Menanteau05}. 
More important, cosmic variance in likely to overwhelm 
the systematic normalization issues above for the small 
volumes probed in the observations shown. The difficulty of identifying mergers 
without deep imaging has naturally limited the effective volume of 
most such measurements. \citet{Somerville04b} and \citet{Wolf05} 
discuss this for several of the fields we have considered 
(e.g.,\ GOODS, GEMS, HDF), and for the specific application 
to different morphological classifications, and they estimate 
a factor of $\sim1.5-2$ uncertainty in their estimates from 
GOODS. Lacking a sufficiently large volume to overcome
cosmic variance, we should ideally compare measurements from the 
same fields.  That is, regardless of cosmic variance in a particular field, 
the observed QLF, MGLF, and 
merging galaxy SFRFs in that field should be consistent. 
In fact, if we limit our comparisons 
in this manner, we find very good agreement between the observations and our 
predictions (see e.g.\ \S~\ref{sec:comparisons}). 
The problem here is that again, existing
fields with deep space-based imaging are too small to contain
rare, luminous quasars.
In general, our predictions from 
the QLF may be more representative of the cosmic mean, as they agree 
with the observed QLFs from large volume surveys such as the SDSS 
\citep[e.g.,][]{Richards05,Richards06}.

Despite systematic uncertainties, the trend in $\rho({\rm SFR})$ appears 
robust, and we can make several comparisons between the SFR density in mergers and the
total SFR density estimated by \citet{HopkinsBeacom06}.
Uncertainties in this comparison are dominated by the
factor $\sim2-3$ scatter in observational estimates of the total 
SFR density, rather than uncertainties in our predictions.
Regardless, we confirm the results of various observational
estimates of merger fractions and their evolution with redshift
\citep{LeFevre00,Patton02,Conselice03,Lin04,Bundy04,Conselice05,Bell05,Wolf05}, 
namely that at $z\lesssim1$, star formation triggered in mergers
contributes only a small fraction to the SFR density of the Universe,
but this increases near the era of peak merger and quasar
activity at $z\gtrsim2$. Specifically, from Figure~\ref{fig:sfr.int}
we predict that only $\sim1\%$ of the SFR density at $z=0$ is driven by mergers,
and even assuming a maximal $\fgas=1.0$ in all mergers increases
this only to $\sim10-20\%$ (setting a strong upper limit).  The predicted
contribution from mergers rises to $\sim15-25\%$ by $z=1$, with an upper
limit $\sim35\%$ ($\fgas=1$).  At $z=2$, the contribution from mergers
increases to $40^{+8}_{-6}\%$ of the cumulative SFR density, and
$35^{+10}_{-8}\%$ at $z=3$. Above these redshifts, the best-fit
prediction approximately preserves these ratios, but the uncertainties
become large. Thus, spiral galaxies (and increasingly 
irregular galaxies at higher redshift, e.g.\
\citet{Cross04,Daddi04,Somerville04a}) dominate the 
SFR density at $z\lesssim2$, even assuming the most
optimistic $\fgas=1.0$ pure gaseous mergers, with the contribution 
from mergers rising to a comparable but not dominant $\sim30-40\%$ 
at higher redshifts. 

These predictions do depend systematically on the faint-end slope of
the MGLF, or equivalently the low-mass slope of the merger mass
function. We have adopted typical measured values from
\citet{Xu04,Bundy05a,Bundy05b}, but if the low-mass incompleteness is
large and we instead adopt the maximally steep slope from
\citet{Wolf05}, this increases our predictions by
$\sim50\%$, enhancing the merger-driven contribution to $\rho({\rm
SFR})$ to $\sim30\%$ at $z=1$ and $\sim60\%$ at $z=2$. Moreover,
$\fgas$ is somewhat uncertain, but it enters approximately linearly in
Figure~\ref{fig:sfr.int}, so our prediction can be rescaled with
improved observational constraints on $\fgas$.

By convolving our predicted rate of spheroid formation as a function
of redshift with the characteristic gas fraction as a function of
redshift, we can estimate a ``mean'' $\fgas$ for spheroid-producing
mergers leading to spheroids of a given $z=0$ mass.  The
redshift-dependent rate of spheroid formation is given in
\citet{H05e}, but is essentially determined by the merger rates as a
function of total, final mass at each redshift. For example, the
$\fgas=1$ line in Figure~\ref{fig:sfr.int} can also be thought of as
the total rate at which stellar mass is ``moved'' from the blue
sequence to the red sequence by quasar-producing, gas-rich mergers.
Because the implied merger activity peaks at $z\gtrsim2$ corresponding
to the peak of quasar activity, assuming $\fgas=10^{-t/t_{H}}$ gives a
nearly constant characteristic mean gas fraction $\sim60\%$ for
essentially all spheroid masses of interest
($M_{\ast}\gtrsim10^{8}\,M_{\sun}$).  The old ages of spheroids
coupled with the expected increase in disk gas fractions at high
redshift means that it is indeed a good approximation to assume that
most spheroids are formed in quite gas-rich mergers. This is
supported by observations, as e.g.\ \citet{Hoekstra05} find
from stellar population analysis that elliptical galaxies must have
formed much of their stellar mass around the time of the
elliptical-producing event.  This does not, of course, prohibit
subsequent spheroid-spheroid mergers, which will not trigger star
formation or quasar activity.

\section{Quasar Host Galaxy Luminosity Functions}
\label{sec:HGLF}

Given the luminosity function of merging galaxies which produce quasars, 
it is relatively simple to convert this into the
expected luminosity function of quasar ``host galaxies'' (HGLF), but 
we must first define what we mean by
quasar ``hosts.'' If we include e.g.\ starbursts with buried, low-luminosity 
X-ray AGN then we will recover a similar merger
luminosity function to that we calculate from the X-ray QLF. However,
most efforts to measure the distribution of AGN host galaxy
luminosities
\citep[e.g.,][]{Bahcall97,McLure99,Falomo01,Hamilton02,Jahnke03,Dunlop03,Floyd04,VandenBerk05} 
have considered hosts of bright, {\em optical}, broad-line
quasars.  In our modeling, this phase of quasar activity is associated
with the ``blowout,'' i.e.\ the final stages of black hole growth,
when the surrounding gas is expelled and heated and the black hole is
briefly rendered a bright optical source before accretion shuts
down. We consider this in detail in \citet{H05e}
(Equation~[26]) and derive the useful mean relation from our simulations 
\begin{equation}
\frac{L_{\rm B,\ obs}^{\rm qso}}{L_{\rm B,\ obs}^{\rm gal}}=7.9\,\frac{1}{\fgas}\,
{\Bigl(}\frac{M_{\rm BH}^{f}}{10^{8}\,M_{\sun}}{\Bigr)}^{0.2}\ \frac{L}{\Lp} \, .
\label{eqn:BL.scaling}
\end{equation}
We can use this to estimate the optical luminosity function
of quasar hosts by taking our already determined optical/UV merging
galaxy luminosity functions and adding the selection
criteria appropriate to these galaxies being identified as the hosts
of bright optical quasars.

We show in \S~\ref{sec:guts.lum} that the median merger 
optical/UV luminosity is roughly constant in time, but the quasar 
luminosity can vary by orders of magnitude and spends 
much of its time at luminosities well below $\Lp\sim L_{\rm Edd}(M_{\rm BH})$. 
Consequently, the observed quasar luminosity
is uncorrelated with the host galaxy luminosity. This is discussed
further in \citet{H05c,H05e} in the context of the observed lack of
correlation between quasar luminosity and black hole mass
\citep[e.g.,][]{Ho02,Heckman04,Hao05}, but similarly
explains the observed lack of a correlation between nuclear and host
galaxy optical luminosity
\citep{Bahcall97,McLure99,Jahnke03,Hao05,VandenBerk05}. 
At the brightest luminosities above the break in the QLF, 
this is no longer strictly true, as black holes do
increasingly tend to be at a high Eddington ratio and near their peak
luminosity, and a correlation between nuclear and host luminosity is
expected (although at these luminosities measuring host properties is 
most difficult). 

%\clearpage
\begin{figure}
    \scalesinglefig
    \centering
    \plotone{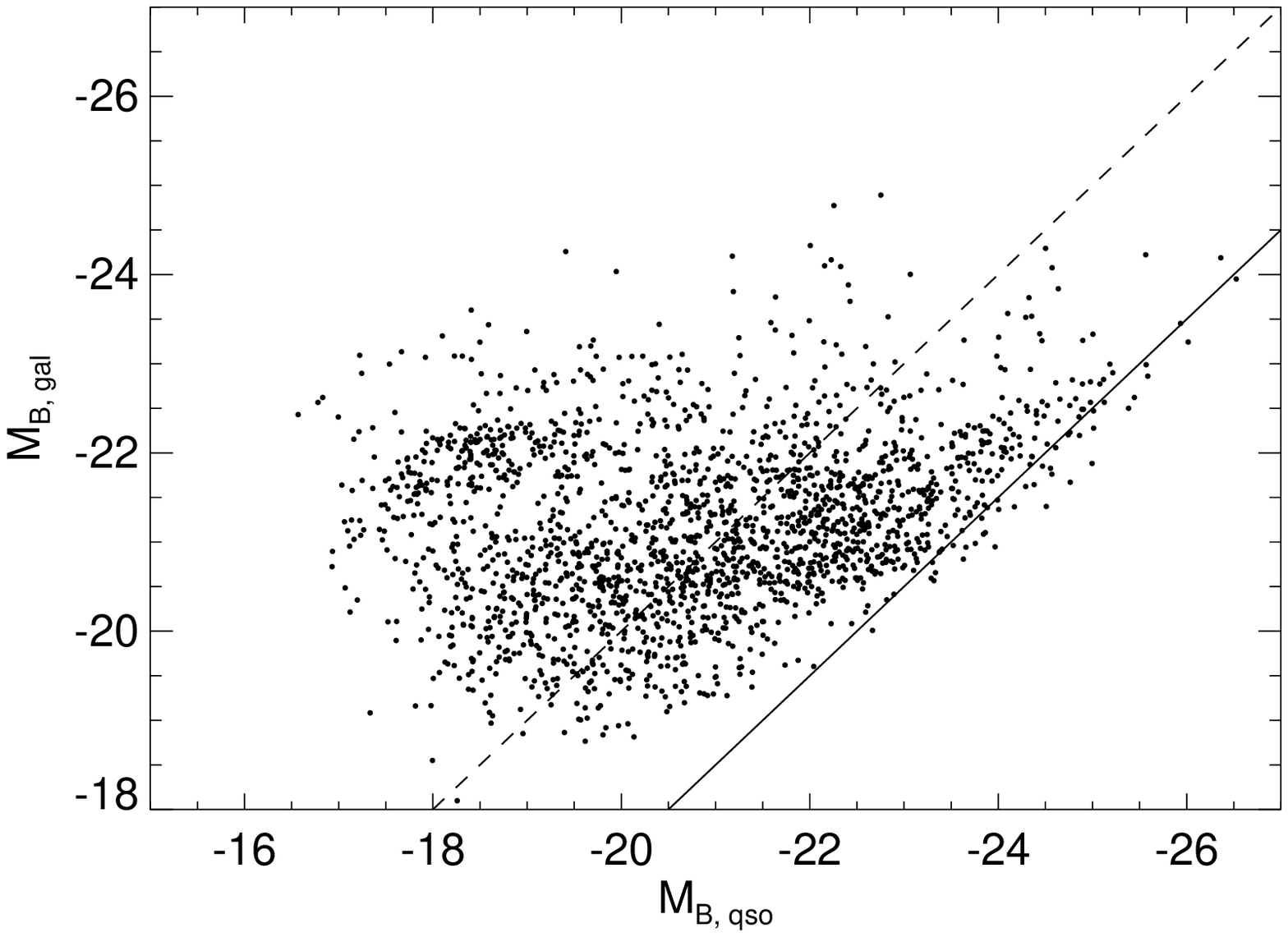}
    \caption{Predicted optical $B$-band quasar and host galaxy magnitudes for a 
    mock sample of $\sim2000$ optical ``AGN'' with $\Lp>10^{9}\,L_{\sun}$, 
    at redshift $z=0.5$. Dashed line shows $L_{B,\ \rm gal}=L_{B,\ \rm qso}$ 
    ($L_{\rm QSO}\approx 0.1\,L_{\rm Edd}$), 
    solid line $L_{B,\ \rm gal}=0.1\,L_{B,\ \rm qso}$ 
    ($L_{\rm QSO}\approx L_{\rm Edd}$). Quasar and host luminosities 
    are uncorrelated except at the brightest luminosities where systems are 
    generally near-Eddington, as observed \citep[e.g.,][]{Hao05}.
    \label{fig:gal.qso.LB}}
\end{figure}
%\clearpage

Using our modeling of the quasar and galaxy
luminosities during a merger, we can predict the joint distribution in
optical quasar and host galaxy luminosity. 
In Figure~\ref{fig:gal.qso.LB}, we generate a mock sample of
$\sim2000$ optical ``AGN'' and plot their quasar and host galaxy
$B$-band magnitudes.  The procedure by which the mock distribution is
generated is described in detail in \citet{H05e}, but briefly, we
consider intervals of $\log(M_{\rm BH})$ from the distribution of relic 
black hole masses $\nLp$ at $z=0.5$ with
$10^{6}\,M_{\rm BH}<\Lp<10^{10}\,M_{\sun}$. For each, we consider the
simulations with the nearest $M_{\rm BH}$, and calculate the PDF
for their being observed as optical AGN (defined here by $M_{B,\ \rm
qso}<-16$ and an obscuring column density $N_{H}<10^{22}\,$cm$^{-2}$)
with a given AGN and host galaxy $B$-band luminosity (both calculated
including attenuation using the methodology in \citet{H05e}, but
adopting the empirical fit to column density distributions as a function of 
luminosity from \citet{Ueda03} yields very similar results). From this
distribution, we randomly calculate $\sim2000$ points in $M_{B,\ \rm
qso}$, $M_{B,\ \rm gal}$, which are shown in the figure.  We assume
$\fgas=0.2$, but this only changes the normalization of $M_{B,\ \rm
gal}$ in this distribution.  For comparison, the dashed line in the
figure shows $L_{B,\ \rm gal}=L_{B,\ \rm qso}$ and the solid line
shows $L_{B,\ \rm gal}=0.1\,L_{B,\ \rm qso}$. These 
approximately correspond to Eddington ratios of $L\approx0.1\,L_{\rm Edd}$ and 
$L\approx L_{\rm Edd}$, respectively. The distribution
demonstrates the lack of correlation between quasar and host galaxy
luminosity, and agrees well with various observational estimates
\citep[e.g.,][]{Bahcall97,McLure99,Jahnke03,Sanchez04}. 
Specifically, compare Figure~13 of Vanden Berk et al.\ 2006, who find
a nearly identical distribution considering host galaxy-AGN spectral
decomposition of SDSS AGN complete to $L_{g,\ \rm
gal}\approx0.1\,L_{g,\ \rm qso}$ (where the SDSS $g$-band is generally
equivalent to $B$-band for our purposes).

%\clearpage
\begin{figure}
    \scalesinglefig
    \centering
    \plotone{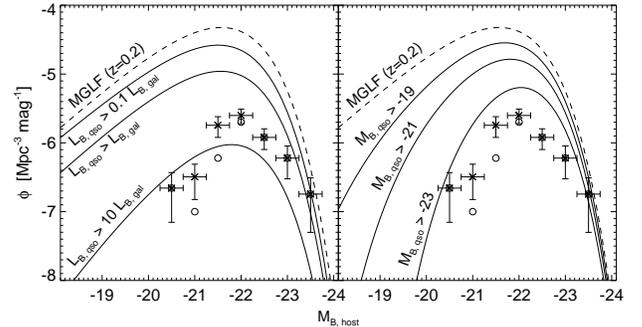}
    \caption{Predicted $B$-band HGLF of optical quasars at $z=0.2$. Dashed 
    line shows the full best-fit $B$-band MGLF (assuming $\alpha=0.5$), 
    solid lines show the HGLF requiring an observed $B$-band quasar 
    luminosity above a given fraction of the host galaxy luminosity (left 
    panel, as labeled) or requiring an observed $B$-band quasar 
    magnitude above some fixed limit (right panel, as labeled). 
    Points show the observational estimate of \citet{Hamilton02}
    (requiring $M_{V,\ \rm qso}<-23$, open circles) at $z\sim0-0.4$, 
    rescaled to typical $B$-band host luminosities following 
    \citet{Bahcall97} and \citet{VandenBerk05}. Although these quasars 
    are not necessarily predicted to be observed in an obvious merger stage, their 
    host luminosity function is a subset of the MGLF, selected at a different merger stage.
    \label{fig:HGLF}}
\end{figure}
%\clearpage
For a given $M_{\rm BH}$ (or
correlated merging galaxy luminosity), we can calculate from
Equation~(\ref{eqn:BL.scaling}) and our scaling of the quasar lifetime
as a function of instantaneous and peak luminosities the time spent
above some observed quasar $B$-band luminosity limit $L_{B,\ \rm qso}$
and/or some limit in the ratio of quasar to host galaxy optical
luminosity $L_{B,\ \rm qso}/L_{B,\ \rm gal}$; i.e.\ the time
during which we estimate the merging galaxy can be identified as a
traditional optical or broad-line AGN host galaxy.
Figure~\ref{fig:HGLF} shows this calculation assuming
different selection criteria in determining the HGLF.  In the left
panel, we show the MGLF at $z=0.2$, determined from the $z=0.2$ hard
X-ray QLF of \citet{Ueda03} as the dashed line. For clarity, we do not
show the uncertainty in this prediction, but adopt the best-fit
assuming $\alpha=-0.5$ and $\fgas=0.2$; the
uncertainty in the plotted MGLF can be seen in the upper and lower
middle panels of Figure~\ref{fig:LFs}.  As solid lines, we then show
the HGLF, i.e.\ the MGLF given some selection criteria for the quasar
and host galaxy $B$-band luminosity, namely $L_{B,\ \rm
qso}>0.1\,L_{B,\ \rm gal}$, $L_{B,\ \rm qso}>L_{B,\ \rm gal}$, and
$L_{B,\ \rm qso}>10\,L_{B,\ \rm gal}$, as labeled.  In the right
panel, we show the same, but instead imposing the $B$-band quasar
magnitude limits $M_{B,\ \rm qso}<-19$, $M_{B,\ \rm qso}<-21$, and
$M_{B,\ \rm qso}<-23$, as labeled. In each, the observationally
estimated optical quasar HGLF of \citet{Hamilton02} is plotted, which
is composed of quasars primarily of optical magnitude $M_{B}<-23$
(crosses with error bars; open circles show the inferred luminosity
function applying a strict $M_{V}<-23$ magnitude cut). 
There is some uncertainty in the absolute normalization of the 
host galaxy luminosities, discussed in \citet{Hamilton02}; 
therefore we rescale their $V$-band host galaxy luminosities 
to the $B$-band using the calibrations of the overlapping samples in 
\citet{Bahcall97,McLure99}, which gives a peak in the HGLF 
similar to that observed in e.g.\ \citet{Bahcall97}
and the much larger sample of \citet{VandenBerk05} at $M_{B}\approx-22$.

In any case, when we require that a bright optical 
quasar ($M_{B}<-23$) be observable as such, we find
good agreement between our predicted HGLF and the observed
distribution of host galaxy luminosities.
There are several
uncertainties in this comparison: for example, the normalization and
absolute host magnitudes of the observed sample may be biased 
if the quasar light is not perfectly subtracted, our
prediction only loosely estimates the full distribution of dust
obscuration as a function of time (see Hopkins et al.\ 2006b for a
discussion; the key point is that by $M_{B}<-23$, a more accurate
prediction from our simulations of the HGLF may be up to a factor
$\sim2$ lower), and as shown in Figure~\ref{fig:LFs} the uncertainties
in the initial MGLF (which propagate linearly into the HGLF) 
are large.  For these reasons, the slight
discrepancy in normalization between our prediction and the
observations is not significant. Also, note that the appearance in
Figure~\ref{fig:HGLF} that all bright mergers host optical quasars is an
artifact of how steeply the Schechter function falls off -- even at
the brightest luminosities, there can be a factor $\sim2-10$
difference between the MGLF and $M_{B}<-23$ HGLF prediction.

However, despite these uncertainties, the key qualitative point is
clear -- the quasar HGLF can be understood as a subset of the MGLF,
with the appropriate though more sophisticated selection criteria
applied as only those mergers during (or after) the ``blowout''
phase will be identified.  The characteristic magnitude, relative
normalization, and narrow width of the HGLF are explained in our
modeling, demonstrating that this luminosity function is
self-consistent with both the QLF and MGLF in our interpretation of
quasar fueling. Observations of e.g.\ the 
fraction of mergers with optical AGN
\citep[e.g.,][]{Sanchez05}, 
although traditionally more difficult than observations of the
fraction of AGN in mergers, also suggest a similar fraction of mergers
hosting optical AGN to that we predict for the appropriate luminosity
limits in Figure~\ref{fig:HGLF}.  Although we have only shown this at
one redshift, where most of the observations have been made, the
qualitative result is the same regardless of the redshift (or e.g.\
value of $\fgas$ or $\alpha$) chosen.

This does not imply that all quasar host galaxies will
be visible as mergers or interacting systems, although the
low-redshift sample of \citet{Bahcall97} indeed identifies the vast
majority as such.  Generally, the ``blowout'' phase in our simulations
follows the final coalescence of the black holes, meaning that tidal
tails and other evidence of a recent merger should in principle be
evident, but at higher redshift surface brightness dimming will make
these features nearly impossible to observe. Moreover, once the blowout
begins, star formation is terminated and the host galaxy rapidly
begins to redden, and extended tidal features will fade in
$\sim 1$ Gyr.

The HGLF distribution we predict is consistent with
observed properties of lower-luminosity AGN host galaxies. For
example, \citet{Kauffmann03} study the host galaxies of $\sim20,000$
SDSS narrow line, relatively low-luminosity AGN at $z<0.3$ and find
that a large fraction of these objects correspond to what we expect
from simulations for post-``blowout'' objects with rapidly
declining accretion rates in relaxing, rapidly reddening systems (see
e.g.\ Hopkins et al.\ [2006a] for a description of the falloff in AGN
luminosity as the merger relaxes). These objects reside in massive
spheroids, with properties of ``normal'' ellipticals except for young
stellar populations and evidence of starbursts in the past
$\sim1-2\,$Gyr, with a sizable fraction ($\sim30\%$) of especially
the brightest objects showing obvious evidence of interaction and/or
recent mergers. Similar results are also found for e.g.\ the AGN host
population in GEMS \citep{Sanchez04} and radio loud quasars
\citep{Sanchez03}, although these may be preferentially at relatively
low accretion rates \citep[e.g.,][]{Ho02,HNH06}.

\section{Discussion}
\label{sec:discuss}

We have used simulations of galaxy mergers which account for star
formation, metal enrichment, radiative cooling, supernova feedback and
pressurization of a multi-phase interstellar medium, and black hole
growth and feedback, to relate the distribution of observed quasar
properties and its evolution with redshift to the distributions of
merging galaxy properties, including their luminosity and mass
functions, characteristic gas fractions, and contribution to the star
formation rate density of the Universe.  Our simulations span a wide
range of initial and final conditions, varying virial velocities,
initial gas fractions, masses, orbital parameters, initial black hole
masses, ISM gas equations of state, redshifts, and galaxy mass ratios.
Our modeling allows us to self-consistently map between the merging
galaxy and quasar distributions in a physically motivated manner
without invoking tunable cosmological distributions and enables
predictions of many properties of merging galaxies at different
redshifts.

We find that:

{$\bullet$ }{The joint scaling of dust obscuration and star formation
with gas density yields quite flat lightcurves in the optical/UV, allowing 
us to provide a simple parameterization of the probability for observing a 
particular
merger at a certain luminosity. To enable future comparison with observations, 
we provide fits
from our simulations to infer the distribution of observed
luminosities in different bands as a function of the stellar mass
formed in mergers, total stellar mass, or final black hole mass of the
merging systems (Table~\ref{tbl:magslookup}). Our simulations allow us to 
determine color and mass-to-light ratio distributions of mergers, which we
find to be
in good agreement with those observed, and self-consistently map between 
e.g.\ observed merger luminosity functions in the near IR, optical, and UV, 
star formation rate distributions, mass functions, and mass or luminosity-dependent 
merger rates.}

{$\bullet$ }{We consider in detail observed merger
luminosity functions in $K$-band at $z<0.2$ \citep{Xu04} and  
280\,nm at $z=0.7$ \citep{Wolf05}, and mass functions 
at $z\sim0.4,\ 0.7,\ 1.2$ \citep{Bundy05a,Bundy05b}. We use our
simulations to map them to a quasar luminosity function at each 
redshift, and find that
the predicted quasar luminosity functions agree with those observed by e.g.\ 
\citet{Ueda03}. Conversely, we invert this
procedure, predicting the merger luminosity function from the observed
quasar luminosity function, and again find agreement, although the
predictions are significantly less well-constrained in this direction.
Both distributions are 
self-consistent (each predicts the other) and can be used to predict
merger rates and the rate of formation of quasars or spheroids 
as a function of that luminosity, mass and redshift.}

{Although merger mass functions are not yet
well determined above these redshifts, the demonstration in \citet{H06} 
that the characteristic masses of observed mergers, the 
``quenching'' or ``transition'' mass where elliptical galaxies begin to 
dominate the total galaxy population, and the characteristic mass of 
quasars (at the observed QLF break) all trace one another 
over the range $0<z\lesssim3$ further suggests that this 
mapping is valid at high redshift.  We extend our predictions of 
merger luminosity functions by using the observed QLF to 
predict the merger luminosity density in $K$ and $B$-bands from 
$z=0-6$ with small factors $\sim2$ uncertainties. The prediction agrees with 
existing observations at $z\lesssim3$ \citep{Brinchmann98,Toledo99,
Conselice03,Conselice05,Xu04,Bundy05a,Bundy05b}, 
but the present observational 
uncertainties are large. Future measurements of the mass or 
luminosity density in mergers will provide a critical test of this model
and the association between mergers and quasar triggering at all redshifts.}

{$\bullet$}{We compare this relation of merger and quasar distributions to that
inferred if quasars turn ``on/off'' in step-function fashion or
follow exponential lightcurves, and find that the observations rule
out such scenarios at $\gtrsim99.9\%$ confidence, even if we allow the
quasar lifetime to vary freely to produce the best fit. This can be
determined entirely from the observed {\em shapes} of the QLF and
MGLF, independent of selection efficiencies, gas fractions, or other
cosmological properties; it would require a $\gtrsim4-6\sigma$ 
change in the observed shape to reverse this conclusion. Any
comparison of merger rates or merger distributions and quasar activity
must account for complex quasar lightcurves and the non-trivial
nature of quasar lifetimes which vary as a function of both
instantaneous and peak luminosity.}

{$\bullet$}{We use the QLF to predict the distribution of 
star formation rates (SFRF) at various redshifts, and
subsequently the SFR density in mergers as a function of redshift from
$z=0-6$.  Comparison of these predictions with measurements, which
currently exist at $z\lesssim1$
\citep{Brinchmann98,Menanteau01,Menanteau05,Bell05}, shows good
agreement, and suggests that the QLF can provide a new, powerful, and
independent probe of the SFR density in the Universe driven by
mergers.  At redshifts $z\lesssim4$, this allows us to predict the SFR
density caused by mergers to a factor $\sim1.5-2$ accuracy, with the
primary systematic uncertainty being weak constraints on typical gas
fractions of disks at low redshift $z\lesssim1$. At $z>4$, future
improvements in constraints on the QLF break luminosity will enable
similar predictions, without a large uncertainty in gas fractions, as
gas-rich systems are expected at these high redshifts. Thus, quasar
measurements can constrain the SFR density
triggered by mergers at high redshifts and at the end of the
reionization epoch, where attempts to measure morphological properties
of galaxies and determine this quantity directly are not likely to be
possible in the near future.  This also provides a prediction of the
total rate at which mass is moved from the blue (disk) sequence to the
red (spheroid) sequence, independent of gas fraction, as analyzed in
detail in \citet{H05f}.}

{We find that the SFR density triggered in mergers is small at low
redshift: $\sim1\%$ of the total SFR density at $z\sim0$. 
This fraction of the SFR
density rises to a substantial, but not dominant $\sim40\%$ (upper
limit $60\%$) at $z\sim2-3$, and remains approximately constant at
higher redshift (relative to the total 
SFR density from \citet{HopkinsBeacom06}). Given 
the low level of star formation in remnant
ellipticals, the dominant contribution to the SFR
density (especially at $z\lesssim2$) must come from
late-type spiral or irregular galaxies.}

{$\bullet$ }{We predict the joint distribution of quasar and host
galaxy luminosities in good agreement with observations
\citep{Bahcall97,McLure99,VandenBerk05}.  The lack of a correlation
between quasar and host galaxy luminosities
\citep[especially at low quasar luminosity;][]
{Bahcall97,McLure99,Ho02,Jahnke03,Heckman04,Hao05,VandenBerk05} is a
natural consequence of the nearly constant optical/UV merger light
curves we discuss above while quasar luminosities change by orders of
magnitude during a merger.}

{We predict the quasar host galaxy luminosity function (HGLF) from either the
observed QLF or merging galaxy luminosity function 
(MGLF) at any given redshift.  We find that the HGLF has
a similar shape to the MGLF in a given band, but becomes increasingly
peaked about $M_{\ast}$ as the minimum optical quasar luminosity or
minimum ratio of optical quasar to optical host galaxy luminosity is
increased.  For a given minimum quasar optical luminosity, e.g.\
$M_{B}<-23$, we find good agreement between the observed HGLF
\citep{Bahcall97,Hamilton02,VandenBerk05} and our prediction from the
observed MGLF of \citet{Xu04} or QLF of \citet{Ueda03}.  The observed
shape, normalization, and break/turnover luminosity in the quasar HGLF
is a natural consequence of its being a subset of the MGLF, with e.g.\
a break at approximately the same luminosity (so long as the
luminosity limit of the quasars in the sample is not so high as to
exclude typical $\sim M_{\ast}$ hosts). This does not mean
that quasar hosts will be obvious mergers, as the
bright optical phase of quasar activity is associated with the
``blowout'' phase following the final coalescence, in which difficult to 
observe, rapidly
fading tidal features may represent the only morphological merger
signature.}

We have demonstrated that the statistics and distributions of merging
galaxy and quasar populations are self-consistent and can be used to
predict one another in the context of the merger hypothesis. Coupled
with the analysis of quasar properties in Hopkins et al.\ (2005a-d, 2006a,b), 
remnant red-sequence elliptical galaxies in Hopkins et al.\
(2006c), and the co-evolution of the ``transition'' mass and merger and 
quasar masses in \citet{H06}, the modeling presented here unifies the populations of
multiple relevant merger stages. The number of observed mergers
accounts for the bright quasar population as well as the observed
buildup in the mass of the elliptical / red-sequence galaxy
population, and vice versa. {\em There is no room for a large fraction of
gas-rich mergers which do not produce a bright quasar phase and
remnant, reddening elliptical, nor is there room for a large fraction
of bright quasars or elliptical galaxies which are not formed in
gas-rich mergers.} 

This allows us to make a wide range of predictions
and provides a critical test of the hypothesis that starbursts,
quasars, and elliptical galaxies are linked through the process of
gas-rich mergers, a test which is complimentary and equally important
to measurements of the individual photometric and kinematic properties
and correlations of mergers and merger remnants \citep[e.g., their
profiles, metal enrichment, phase space densities, fundamental plane
and $M_{\rm BH}-\sigma$ relations: ][]
{LakeDressler86,Doyon94,Oliva95,
ShierFischer98,James99,Genzel01,RJ04,RJ05} and comparison of these
remnants to detailed simulations
\citep[e.g.,][]{BH92,H93a,
Hernquist93,SDH05a,Cox05,Robertson05a,Robertson05b}.

This does not mean, of course, that ellipticals never evolve via dry
(gas-poor, spheroid-spheroid) mergers \citep[see, e.g.,][]{Bell05b,vanDokkum05}, 
as such mergers only modify a
necessarily pre-existing population of ellipticals (see also
\citet{H05f} for a calculation of the effects of such processes on
observed elliptical distributions). This also does not imply that all
AGN activity is associated with gas-rich mergers, as there can be a
substantial contribution to low-luminosity AGN from activity triggered
in quiescent spirals \citep[for a detailed comparison, see][]{HH06}, 
as well as the large amount of quiescent,
low-level activity in relaxed ellipticals which we predict (and
account for) many dynamical times after the spheroid-forming merger as
the quasar lightcurve decays \citep{H05e}.

Our modeling makes predictions that can be used to test our underlying
theory.  For example, mergers at high redshifts should involve
galaxies that are, on average, more gas-rich than local spirals 
(our preliminary comparison, for example, favors an exponentially 
declining $\fgas$ over $\sim$\,a few Gyr).
Preliminary evidence from \citet{Erb06}
indicates that galaxies at redshift $z\approx 2$ do, indeed, have
large gas fractions $\fgas\sim0.5$, with some approaching
$\fgas\sim0.8 - 0.9$, but future observatories such as ALMA should be able to 
measure this accurately. The connection between merger-driven
star formation and quasar activity can also be used to test the 
correlation between our predicted SFR distributions and quasar 
peak luminosity. This should be possible with improved observations of 
quasar hosts in relaxed, post-merger systems which can be used to 
determine their individual star formation histories \citep[e.g.,][]{Kauffmann03}, as 
well as measurements of star formation in obscured, lower-luminosity 
merger phases, possibly associated with IR-bright Type 2 quasar 
activity \citep[e.g.,][]{Hao05b,Kim06} and strongly reddened optical quasars 
\citep[e.g.,][]{Urrutia05}.

Mergers of gas-rich spirals will imprint structure into the remnants
that may be difficult to account for otherwise 
\citep[see e.g.,][]{Robertson05b,Cox05}.  The starburst
population left behind will characteristically modify the central
light profiles and kinematics of merger remnants 
(e.g.\ Mihos \& Hernquist 1994b; Hernquist \& Barnes 1991),
possibly explaining the central excesses of light seen in merging
systems (e.g.\ Rothberg \& Joseph 2004, 2006), and the presence of
kinematic subsystems in ellipticals.
This can be tested in
detail by comparing predictions for metallicity and color gradients
and orbital distributions with observations. 

More subtle may be the shells, ripples, loops
and other fine structures seen around many relaxed ellipticals (see,
e.g.\ Schweizer 1998) that require a source of dynamically cold
stellar material and are hence a natural consequence of major mergers
involving disk galaxies (e.g.\ Hernquist \& Spergel 1992), but do
{\it not} form in major mergers between hot stellar systems
(e.g. Hernquist \& Quinn 1988).  A measurement
of the rate of occurrence of fine structure in red galaxies with
redshift would further constrain the importance of mergers involving
disk galaxies to the formation of ellipticals \citep{Bell05b,vanDokkum05}.

We use our simulations to develop a formalism to derive the relations
between the various populations we have studied in a manner robust
against different cosmological distributions which are poorly
constrained and often tuned to reproduce observations. However, this
also means that we cannot constrain certain cosmological
distributions.  For example, we find that the distribution of observed
luminosities of a merger is robust when expressed as a function of the
final stellar mass or total stellar mass formed in the merger. This
enables us to map the MGLF to a merger mass function, but does not
allow us to consider the relative contribution of mergers with e.g.\
different initial mass ratios, ISM equations of state, or star
formation histories. It will be interesting to see whether cosmological simulations 
yield the distributions of merger statistics that we have 
derived from quasar, merger, and elliptical galaxy observations. Coupling such
cosmological descriptions with our detailed modeling of the complex
star formation and black hole growth histories in mergers enables an a
priori theoretical prediction of the wide array of phenomena we have
demonstrated are linked through gas-rich galaxy mergers.

\acknowledgments We thank E. Bell and C. Wolf for providing their data
in electronic form.  This work was supported in part by NSF grants ACI
96-19019, AST 00-71019, AST 02-06299, and AST 03-07690, and NASA ATP
grants NAG5-12140, NAG5-13292, and NAG5-13381.  The simulations were
performed at the Center for Parallel Astrophysical Computing at the
Harvard-Smithsonian Center for Astrophysics. RSS thanks the Institute
for Theory and Computation at the Harvard-Smithsonian Center for
Astrophysics for hospitality.

\clearpage
\begin{landscape}
\begin{deluxetable}{cccccccccccccc}
%\rotate
%\tablecolumns{13}
\tabletypesize{\scriptsize}
\tablecaption{Best-Fit Observed Merger Luminosities\label{tbl:magslookup}}
\tablewidth{0pt}
\tablehead{
\multicolumn{1}{c}{} & 
\multicolumn{3}{c}{\tablenotemark{1}${\langle}M_{\rm BAND}{\rangle}=M_{10}+\alpha\,\log{\Bigl(}\frac{\mnew}{10^{10}\,M_{\sun}}{\Bigr)}$} & 
\multicolumn{1}{c}{} & 
\multicolumn{3}{c}{\tablenotemark{2}${\langle}M_{\rm BAND}{\rangle}=M_{11}+\alpha\,\log{\Bigl(}\frac{\mtot}{10^{11}\,M_{\sun}}{\Bigr)}$} &
\multicolumn{1}{c}{} & 
\multicolumn{3}{c}{\tablenotemark{3}${\langle}M_{\rm BAND}{\rangle}=M_{8}+\alpha\,\log{\Bigl(}\frac{M_{\rm BH}}{10^{8}\,M_{\sun}}{\Bigr)}$} \\
\cline{2-4}\cline{6-8}\cline{10-12}\\
\colhead{Band} &
\colhead{$M_{10}$} &
\colhead{$\alpha$} &
\colhead{$\reducechi$} &
\colhead{} &
\colhead{$M_{11}$} &
\colhead{$\alpha$} &
\colhead{$\reducechi$} &
\colhead{} &
\colhead{$M_{8}$} &
\colhead{$\alpha$} &
\colhead{$\reducechi$} & 
\colhead{} & 
\colhead{$\sigma$}
}
\startdata
%Bolometric & $-00.0\pm0.0$ & $-0.0\pm0.0$ & 0.00 & &
%              $-00.0\pm0.0$ & $-0.0\pm0.0$ & 0.00 & &
%              $-00.0\pm0.0$ & $-0.0\pm0.0$ & 0.00 & & $0.00\pm0.00$ \\
280/40 & $-20.7\pm0.1$ & $-2.0\pm0.1$ & 0.61 & &
              $-22.0\pm0.2$ & $-2.0\pm0.3$ & 2.29 & &
              $-22.3\pm0.2$ & $-1.5\pm0.2$ & 2.76 & & $0.62\pm0.06$ \\
U        & $-19.7\pm0.2$ & $-2.0\pm0.3$ & 0.76 & & 
              $-20.9\pm0.2$ & $-2.3\pm0.3$ & 1.47 & & 
              $-21.3\pm0.2$ & $-1.9\pm0.2$ & 1.59 & & $0.67\pm0.32$ \\
B        & $-20.2\pm0.2$ & $-2.0\pm0.2$ & 0.97 & & 
              $-21.2\pm0.1$ & $-2.3\pm0.2$ & 0.83 & & 
              $-21.7\pm0.1$ & $-1.7\pm0.2$ & 0.77 & & $0.49\pm0.25$ \\
V        & $-20.7\pm0.1$ & $-2.0\pm0.2$ & 1.66 & & 
              $-21.6\pm0.1$ & $-2.2\pm0.2$ & 0.74 & & 
              $-22.1\pm0.1$ & $-1.7\pm0.2$ & 1.13 & & $0.34\pm0.28$ \\
\\
\hline \\
R        & $-21.0\pm0.1$ & $-1.9\pm0.2$ & 2.40 & & 
              $-21.8\pm0.1$ & $-2.2\pm0.2$ & 0.88 & & 
              $-22.3\pm0.1$ & $-1.7\pm0.1$ & 1.61 & & $0.52\pm0.28$ \\
I        & $-21.2\pm0.1$ & $-1.9\pm0.1$ & 3.40 & & 
              $-21.9\pm0.1$ & $-2.3\pm0.1$ & 0.97 & & 
              $-22.5\pm0.1$ & $-1.7\pm0.1$ & 2.24 & & $0.45\pm0.26$ \\
J        & $-22.2\pm0.1$ & $-2.0\pm0.1$ & 6.40 & & 
              $-22.9\pm0.1$ & $-2.2\pm0.1$ & 1.51 & & 
              $-23.5\pm0.1$ & $-1.7\pm0.1$ & 3.69 & & $0.47\pm0.18$ \\
H        & $-22.6\pm0.1$ & $-2.1\pm0.1$ & 6.15 & & 
              $-23.4\pm0.1$ & $-2.2\pm0.1$ & 1.67 & & 
              $-24.1\pm0.1$ & $-1.7\pm0.1$ & 3.96 & & $0.39\pm0.17$ \\
K        & $-22.7\pm0.1$ & $-2.1\pm0.1$ & 5.18 & & 
              $-23.6\pm0.1$ & $-2.3\pm0.1$ & 1.47 & & 
              $-24.3\pm0.1$ & $-1.7\pm0.1$ & 3.42 & & $0.29\pm0.21$ \\
\\
\hline \\
SDSS $u$ & $-19.3\pm0.2$ & $-2.2\pm0.3$ & 0.85 & & 
              $-20.7\pm0.2$ & $-2.2\pm0.3$ & 1.29 & & 
              $-21.1\pm0.1$ & $-1.7\pm0.2$ & 0.84 & & $0.76\pm0.41$ \\
SDSS $g$ & $-20.5\pm0.2$ & $-2.0\pm0.2$ & 1.15 & & 
              $-21.5\pm0.1$ & $-2.2\pm0.2$ & 0.80 & & 
              $-22.0\pm0.1$ & $-1.7\pm0.2$ & 0.85 & & $0.52\pm0.25$ \\
SDSS $r$ & $-20.9\pm0.1$ & $-1.9\pm0.2$ & 2.38 & & 
              $-21.7\pm0.1$ & $-2.2\pm0.2$ & 0.83 & & 
              $-22.3\pm0.1$ & $-1.7\pm0.1$ & 1.58 & & $0.52\pm0.27$ \\
SDSS $i$ & $-21.1\pm0.1$ & $-2.0\pm0.1$ & 3.19 & & 
              $-21.9\pm0.1$ & $-2.2\pm0.1$ & 0.88 & & 
              $-22.5\pm0.1$ & $-1.7\pm0.1$ & 2.08 & & $0.48\pm0.26$ \\
SDSS $z$ & $-21.3\pm0.1$ & $-2.0\pm0.1$ & 4.46 & & 
              $-22.0\pm0.1$ & $-2.3\pm0.1$ & 1.04 & & 
              $-22.6\pm0.1$ & $-1.7\pm0.1$ & 2.65 & & $0.48\pm0.22$ \\

\enddata
\tablenotetext{1}{Distribution of observed luminosities of mergers, 
as a function of new stellar mass $\mnew$ formed during the merger.
This assumes the probability of viewing a merger at 
some magnitude in the given band is a Gaussian with median 
$\langle{M_{\rm BAND}}\rangle$ and dispersion $\sigma$ in the band.}
\tablenotetext{2}{Same,
as a function of total, final stellar mass $\mtot$ after the merger.}
\tablenotetext{3}{Same,
as a function of final black hole mass $M_{\rm BH}$ after the merger.}
\end{deluxetable}
\clearpage
\end{landscape}

\end{document}